\documentclass[aps,preprint,nofootinbib,floatfix]{revtex4-1}
\usepackage{graphicx}
\usepackage{amsmath,amssymb}
\usepackage{url}
\usepackage{color}
\usepackage{epstopdf}
\newcommand{\lsim}{\mathrel{\mathop{\kern 0pt \rlap
  {\raise.2ex\hbox{$<$}}}
  \lower.9ex\hbox{\kern-.190em $\sim$}}}
\newcommand{\gsim}{\mathrel{\mathop{\kern 0pt \rlap
  {\raise.2ex\hbox{$>$}}}
  \lower.9ex\hbox{\kern-.190em $\sim$}}}

\begin{document}
\title{Supersymmetric Decays of the $Z'$ Boson}
\author{Chun-Fu Chang$^{1}$, Kingman Cheung$^{2,1}$, and Tzu-Chiang Yuan$^3$}

\affiliation{
$^1$Department of Physics, National Tsing Hua University, 
Hsinchu 300, Taiwan\\
$^2$Division of Quantum Phases \& Devices, School of Physics, 
Konkuk University, Seoul 143-701, Korea \\
$^3$Institute of Physics, Academia Sinica, Nankang, Taipei 11529, Taiwan
}

\date{\today}

\begin{abstract}
  The decay of the $Z'$ boson into supersymmetric particles is
  studied. We investigate how these supersymmetric modes affect the
  current limits from the Tevatron and project the expected
  sensitivities at the LHC.  Employing three representative
  supersymmetric $Z'$ models, namely, $E_6$, $U(1)_{B-L}$, and the
  sequential model, we show that the current limits of the $Z'$ mass
  from the Tevatron could be reduced substantially due to the
  weakening of the branching ratio into leptonic pairs.  The mass
  reach for the $E_6$ $Z'$ bosons is about $1.3-1.5$ TeV at the LHC-7
  (1 fb$^{-1}$), about $2.5 - 2.6$ TeV at the LHC-10 (10 fb$^{-1}$),
  and about $4.2 - 4.3$ TeV at the LHC-14 (100 fb$^{-1}$). A similar
  mass reach for the $U(1)_{B-L}$ $Z'$ is also obtained.  We also
  examine the potential of identifying various supersymmetric decay
  modes of the $Z'$ boson because it may play a crucial role in the
  detailed dynamics of supersymmetry breaking.
\end{abstract}
\maketitle

\section{Introduction}
Existence of extra neutral gauge bosons has been predicted in many extensions
of the standard model (SM) \cite{paul}.  String-inspired models and
grand-unification (GUT) models usually contain a number of extra $U(1)$
symmetries, beyond the hypercharge $U(1)_Y$ of the SM.  The exceptional group
$E_6$ is one of the famous examples of this type \cite{joanne}.  These extra
$U(1)$'s are broken at some intermediate energy scales between the GUT and the
electroweak scales.  Phenomenologically, the most interesting option
is the breaking of these $U(1)$'s at around TeV scales, giving rise to
extra neutral gauge bosons observable 
at the Tevatron and the Large Hadron Collider (LHC).
Recent developments in model buildings also result in new models
that contain extra gauge bosons. For example, 
little Higgs models \cite{lh} with
additional gauge groups predict a number of new gauge bosons;
the SM gauge bosons propagating in the extra dimensions 
after compactification can give rise to Kaluza-Klein towers of gauge bosons
\cite{extra};  Stueckelberg $Z'$ model connecting a hidden sector to
the visible sector in the context of dark matter \cite{stucek}, 
just to name a few.
We will denote these neutral extra gauge bosons generically by $Z'$
for the general discussion in this introduction section.

Collider experiments such as CDF \cite{cdf} and D\O\ \cite{d0} 
at the Tevatron have
been searching for the neutral extra gauge bosons $Z'$, mainly through
its leptonic decay modes.  The leptonic mode is a very clean channel
to probe for $Z'$ since it may give rise to a
discernible peak above the Drell-Yan background right at the $Z'$ mass,
provided that the size of the coupling strength to SM quarks and leptons 
is not too small. 
Currently, the best limit comes from the negative search at the
Tevatron. The lower mass bound on $Z'$ is about $800-900$ GeV for a
number of $Z'$ bosons of the $E_6$ type and a stronger bound of 
almost $1050$ GeV for the sequential $Z'$, which has exactly the 
same coupling strength and
chiral couplings as the SM $Z$ boson that can be served as a bench mark.
The LHC with just an integrated luminosity of about $40$ pb$^{-1}$
has already set limits on $Z'$s \cite{atlas} almost as good as 
those from the Tevatron. With more luminosity accumulated in the current LHC run
the limit on $Z'$ will improve substantially in the near future.

Most of previous studies on $Z'$ bosons focused on the decays into SM
fermions, and the corresponding limits were obtained based on the
decay into leptons. This scenario is not necessarily a must, but just for
simplicity and fewer choices of parameters.  Indeed, when the mass of
$Z'$ is more than a TeV or even larger it has chances of decaying into
other exotic particles that must be included in the model for various theoretical reasons.  
For instance,
in GUT models or in little-Higgs models there are other fermions needed to cancel
the anomalies, the $Z'$ could decay into these exotic fermions if 
their masses are not too heavy.
Another example is the 
minimal supersymmetric standard model (MSSM)
with electroweak-scale SUSY partners and a $Z'$ which could also 
decay into sfermions and Higgsinos. 
In a recently proposed $Z'$-mediated SUSY-breaking model \cite{z-med},
supersymmetry breaking in the hidden sector is communicated by a $Z'$ boson 
to the visible sector.  The low-energy spectrum includes a $Z'$ boson of
a few TeV and light gauginos, such that the $Z'$ can 
also decay into SUSY particles other than the SM fermions.

In this work, we consider a scenario of $Z'$ boson, which arised from 
$U(1)$ symmetry breaking at around TeV scale, in the context of 
weak-scale supersymmetry, in which all SUSY partners are relatively light 
(a few hundred GeV) except for the squarks (may of order $O(1)$ TeV).
Such a $Z'$ may come from breaking of one of the $U(1)$'s in 
$E_6$ \cite{joanne}, $U(1)_{B-L}$ \cite{b-l}, or $U(1)_B$ \cite{pavel} etc.
Once we specify the $U(1)$ charges for the matter superfields and Higgs 
superfields, the couplings of $Z'$ to MSSM particles are determined.  
We study the decays of $Z'$ and its production at the Tevatron and the LHC.
The decays of $Z'$ will be modified when the SUSY particle masses are
only of order a few hundred GeV, which will then affect the leptonic
branching ratio of the $Z'$.  We investigate how much the limits 
from the Tevatron will be affected, because of the reduction in the leptonic
branching ratio.  We also study how much the sensitivity at the LHC
will be reduced when the SUSY decay modes are open for the $Z'$ boson.
Finally, we study the prospect of using the SUSY decay modes of the $Z'$
to search for the $Z'$ boson itself and investigate its properties.  

We note that some related works had appeared in literature, for example 
in Refs.~\cite{kang,wang,moretti}.
However substantial improvements over these 
previous works have been made in this work. These include

\begin{enumerate}
\item
Ref.~\cite{kang} focused on how the presence of supersymmetric and other
exotic particles of $E_6$ models in the $Z'$ decay can affect the $Z'$ boson
discovery at the Tevatron and the LHC.  In our work, we used the most
current updated limit on $\sigma(Z') \times B(Z'\to \mu^+ \mu^-)$
 to put limits
on the mass of various $Z'$ bosons.  We have illustrated the case of
decaying into SM particles only and the case of including both SUSY
and SM particles. We have shown that the $Z'$ mass limits have
to be relaxed by $20-30$ GeV if including SUSY particles in the decay.
Ref.~\cite{kang} was written in 2004 and certainly our paper used the newest
2011 data.  For the LHC sensitivity we worked out the more realistic
energy-luminosity combinations (7 TeV, 10 TeV, 14 TeV). Refs.~\cite{kang,wang}
did not know about the options of 7 and 10 TeV at their time.

\item
Ref.~\cite{wang} focused on how the discovery potential of sleptons 
can be improved
via the decay of $Z'$ into a slepton pair.  They also studied various 
lightest supersymmetric particle (LSP) 
scenarios and used distributions to determine the masses of sleptons, 
gauginos, and the $Z'$ boson.  In our work, in addition to the slepton-pair,
we also studied the $Z'$ decays into a chargino pair and
into a neutralino pair. We have shown clearly that the presence of $Z'$
is visible in the transverse-mass spectrum.  One can therefore measure
the transverse-mass spectrum and determine if there is a $Z'$ boson. 
This spectrum can also be utilized to estimate the mass differences, and couplings
of $Z'$ to sleptons, neutralinos, and charginos.  This can help us
to understand the underlying supersymmetry breaking mechanism.

\item
 In this work, we study $E_6$ models, $U(1)_{B-L}$, and the sequential 
$Z'$ model, while Ref.~\cite{kang} studied only the $E_6$ models and 
Ref.~\cite{wang} studied only the $U(1)_{B-xL}$ model.
\end{enumerate}

The organization of the paper is as follows.  In the next section, we
write down the interactions and briefly describe a few $Z'$ models
and their extensions to include supersymmetry. In Sec. III, we calculate 
the branching ratios of the $Z'$ boson in various models. In Sec. IV,
we show the shift of the limits for the masses of the $Z'$ in various 
models due to opening of supersymmetric particles.  We estimate
the $5\sigma$ discovery reach at the LHC, including the SUSY decay 
modes in Sec. V.  We further discuss in Sec. VI the SUSY decay modes 
of the $Z'$ boson.  We conclude in Sec. VII. 
Feynman rules that are related
to the $Z'$ are collected in the appendix.

\section{$Z'$ Interactions and Representative Models} 

\subsection{$Z'$ Interactions}

Following the notation of Ref.~\cite{paul,lang-luo}, 
the Lagrangian describing the
neutral current gauge interactions of the
standard electroweak $SU(2)_L\times U(1)_Y$ and extra $U(1)$'s is given by
\begin{equation}
\label{Lag_NC}
- {\cal L}_{\rm NC} = e J_{\rm em}^\mu A_\mu + \sum_{\alpha=1}^{n}
g_\alpha J^\mu_\alpha Z^0_{\alpha \mu}\;,
\end{equation}
where $Z^0_1$ is the SM $Z$ boson and $Z^0_\alpha$ with $\alpha\ge 2$ are the
extra $Z$ bosons in the weak-eigenstate basis.
For the present work we only consider one extra $Z_2^0$ 
mixing with the SM $Z^0_1$ boson. 
Thus the second term of 
the Lagrangian in Eq. (\ref{Lag_NC}) can be rewritten as
\begin{equation}
-{\cal L}_{Z^0_1 Z^0_2} = g_1 Z^0_{1\mu} \left[ \sum_f
 \bar \psi_f \gamma^\mu (g_L^{f} P_L + g_R^{f} P_R ) \psi_f \right] +
 g_2 Z^0_{2\mu} \left[ \sum_f
 \bar \psi_f \gamma^\mu ( Q'_{f_L} P_L + Q'_{f_R} P_R) \psi_f
  \right]\;,
\end{equation}
where for both quarks and leptons
\begin{equation}
g_{L,R}^{f} = T_{3L}^f -  x_{\rm w} Q_f \,,
\end{equation}
and $Q'_{f_L,f_R}$ are the chiral charges of fermion $f$ to $Z^0_2$ 
and $P_{L,R} = ( 1 \mp \gamma_5)/2$.
Here $T_{3L}^f$ and $Q_f$ are, respectively, the third component of the weak
isospin and the electric charge of the fermion $f$.
The chiral charges of various $Z'$ models are listed in Tables~\ref{E6} 
and \ref{b-l}.
The overall coupling constant $g_1$ in Eq. (\ref{Lag_NC}) is the 
SM coupling $g/\cos\theta_{\rm w}$, while
in grand unified theories (GUT) $g_2$ is related to $g_1$ by
\begin{equation}
\label{g2g1}
\frac{g_2}{g_1} = \left(\frac{5}{3}\, x_{\rm w} \lambda\right)^{1/2} \simeq
0.62\lambda^{1/2} \,,
\end{equation}
where $x_{\rm w}=\sin^2\theta_{\rm w}$ and $\theta_{\rm w}$ is the weak mixing
angle.
The factor $\lambda$ depends on the symmetry breaking pattern and the fermion
sector of the theory, which is usually of order unity.

The mixing of the weak eigenstates $Z^0_1$ and $Z^0_2$ to form  mass 
eigenstates $Z$ and $Z'$  are parametrized
by a mixing angle $\theta$:
\begin{equation}
\label{mixing}
\left ( \begin{array}{c} Z \\
                         Z'
        \end{array} \right ) = \left( \begin{array}{rr}
                                  \cos\theta & \sin\theta \\
                                 -\sin\theta & \cos\theta
                                      \end{array} \right ) \;
    \left( \begin{array}{c} Z^0_1 \\
                            Z^0_2
            \end{array} \right ) \;.
\end{equation}
The mass of $Z$ is $M_{Z}=91.19$~GeV. 

After substituting  the interactions of the mass eigenstates $Z$ and  
$Z'$ with fermions are 
\begin{equation}
\label{rule}
-{\cal L}_{Z Z' } = \sum_f g_1 \biggl [
  Z_{\mu} \bar \psi_f \gamma^\mu ( l_s^f P_L + r_s^f P_R ) \psi_f 
 +
  Z'_{\mu} \bar \psi_f \gamma^\mu ( l_n^f P_L + r_n^f P_R ) \psi_f  \biggr ]\,,
\end{equation}
where
\begin{eqnarray}
l_s^f = g_L^{f} + \frac{g_2}{g_1} \, \theta \, Q'_{f_L} \,, &\qquad&
r_s^f = g_R^{f} + \frac{g_2}{g_1} \, \theta \, Q'_{f_R} \,, \\
l_n^f = \frac{g_2}{g_1}\, Q'_{f_L} - \theta \, g_L^{f}\,, &\qquad&
r_n^f = \frac{g_2}{g_1}\, Q'_{f_R} - \theta \, g_R^{f}\,.
\end{eqnarray}
Here the subscript ``s'' denotes the observed SM $Z$ boson and ``n''
denoted the new heavy gauge boson $Z'$.
We have used the valid approximation $\cos\theta\approx 1$ 
and $\sin\theta \approx \theta$.  
In the following, we ignore the mixing 
($\theta =0$) such that the precision measurements for the SM $Z$ boson are not
affected, unless stated otherwise. 

\subsection{Including Supersymmetry}

The superpotential $W$ involving the matter and Higgs superfields in a $U(1)'$
extended MSSM can be written as 
\begin{equation}
W = \epsilon_{ab} \left [ y^u_{ij} Q^a_j H_u^b U^c_i 
  - y^d_{ij} Q^a_j H_d^b D^c_i 
  -  y^l_{ij} L^a_j H_d^b E^c_i 
  +  h_s S  H_u^a H_d^b \right ] \;,
\end{equation}
where $\epsilon_{12}= - \,\epsilon_{21} =1$, $i,j$ are family indices,
and $y^u$ and $y^d$ represent the Yukawa matrices for the 
up-type and down-type quarks respectively.
Here $Q, L, U^c, D^c, E^c, H_u$, and $H_d$ denote the MSSM superfields for the 
quark doublet, lepton doublet,
up-type quark singlet, down-type quark singlet, lepton singlet,
up-type Higgs doublet, and down-type Higgs doublet respectively,
and the $S$ is the singlet superfield. Note that we have assumed other
exotic fermions are very heavy. The $U(1)'$ charges of the fields
$H_u, H_d,$ and $S$ are related by $Q'_{H_u} + Q'_{H_d} + Q'_S = 0$ such 
that $S H_u H_d$ is the only term allowed by the $U(1)'$ symmetry 
beyond the MSSM. Once the singlet scalar field $S$ develops a 
VEV, it generates an effective $\mu$ parameter: $\mu_{\rm eff} = 
h_s \langle S \rangle$.  The case is very similar to NMSSM, except we
do not have the cubic term $S^3$. The singlet field will give rise to
a singlet scalar boson and a singlino, which will mix with other particles 
in the Higgs sector and neutralino sector, respectively.  
In this work, we are 
contented with the assumption that the singlet scalar field and
the singlino are heavy enough that it is out of reach at the LHC and
the mixing effects are negligible.  Detailed phenomenological studies
involving the singlet field will be presented in a future work.
Furthermore, we also take the superpartner, dubbed as $Z'$-ino,
of the $Z'$ boson to be heavy.
Phenomenology involving the singlet scalar boson, singlino, and the
$Z'$-ino of various singlet-extended MSSM 
can be found in Refs.~\cite{vernon}.

Below the TeV scale the particle content is the same as the MSSM plus
a $Z'$ boson. Thus, the superpotential and the
soft breaking terms are the same as in MSSM. Extra
couplings of the $Z'$ boson with the MSSM particles
are coming from the gauge interactions of the extra $U(1)$
and the corresponding supersymmetric vertices of Yukawa interactions.

The gauge interactions involving the fermionic and scalar 
components, denoted generically by $\psi$ and $\phi$ respectively,  
of each superfield are
\begin{equation}
  \label{eq1}
 {\cal L} = \bar  \psi \, i \gamma^\mu D_\mu \, \psi 
   + ( D^\mu \phi)^\dagger \, (D_\mu \phi) \;,
\end{equation}
where the covariant derivative is given by 
\begin{equation}
\label{eq2}
  D_\mu = \partial_\mu + i e Q A_\mu + i \frac{g }{\sqrt{2}} (\tau^+ W^+_\mu
  +\tau^- W^-_\mu ) + i g_1 ( T_{3L} - Q x_{\rm w} ) Z_\mu 
        + i g_2  Z'_\mu  Q'  \;.
\end{equation}
Here $e$ is the electromagnetic coupling constant, $Q$ is the electric 
charge, $\tau^\pm$ are the rising and lowering operators on $SU(2)_L$
doublets and $Q'$ is the chiral charges of the $U(1)$ associated with
the $Z'$ boson.  The interactions of $Z'$ with all MSSM fields
go through Eqs.~(\ref{eq1}) and (\ref{eq2}).

Details and conventions are given in the appendix and the
Feynman rules that involve the $Z'$ boson are listed there as well.

\subsection{Representative models}

\subsubsection{The Sequential model}
 The sequential $Z'_{SM}$ model is a reference model of
extra $Z$ bosons.  It has exactly the same chiral charges as the SM $Z$
boson but at a larger mass. The gauge coupling constant is also taken to
be the same as the SM one, i.e., $g_2 = g_1 = g/\cos\theta_{\rm w}$.  Note,
however, that when SUSY modes are open, the $Z'_{\rm SM}$ can also decay into
sfermions, neutralinos, charginos, and Higgs bosons. In general, experimental
constraints on the sequential model are the strongest, because the other 
$E_6$ models, for example, have smaller gauge coupling constant, as in 
Eq.~(\ref{g2g1}).  

\subsubsection{The $E_6$ models}
Two most studied $U(1)$ subgroups in the symmetry breaking chain 
of $E_6$ occur
in
\[
  E_6 \to SO(10) \times U(1)_\psi\,, \qquad 
 SO(10) \to SU(5) \times U(1)_\chi  \;.
\]
In $E_6$ each family of the left-handed fermions is promoted to a 
fundamental $\mathbf{27}$-plet, which decomposes under 
$E_6 \to SO(10) \to SU(5)$ as 
\[
\mathbf{27} \to \mathbf{16} + \mathbf{10} + \mathbf{1} \to
  ( \mathbf{10} + \mathbf{5^*} + \mathbf{1} ) + (\mathbf{5} +
  \mathbf{5^*} ) + \mathbf{1} \;.
\]
Each $\mathbf{27}$ contains the SM fermions, 
two additional singlets 
$\nu^c$ (conjugate of the right-handed neutrino) and $S$, 
a $D$ and $D^c$ pair ($D$ is the exotic color-triplet quark 
with charge $-1/3$ and $D^c$
is the conjugate), and a pair of color-singlet SU(2)-doublet exotics
$H_u$ and $H_d$ with 
hypercharge
$Y_{H_u, H_d} = \pm 1/2$.  In the supersymmetric
version of $E_6$, the scalar components of one $H_{u,d}$ pair can be used
as the two Higgs doublets $H_{u,d}$ of the MSSM. 
The chiral charges $U(1)_\psi$ and $U(1)_\chi$ 
for each member of the $\mathbf{27}$
are listed in the third and fourth columns in Table~\ref{E6}.
In general, the two $U(1)_\psi$ and $U(1)_\chi$ can mix to form
\begin{equation}
  Q'(\theta_{E_6} ) = \cos \theta_{E_6} Q'_\chi + \sin \theta_{E_6} Q'_\psi \;,
\end{equation}
where $0 \le \theta_{E_6} < \pi$ is the mixing angle.  A commonly studied
model is the $Z'_\eta$ model with 
\begin{equation}
 Q'_\eta = \sqrt{ \frac{3}{8} } Q'_\chi - \sqrt{ \frac{5}{8} } Q'_\psi \;,
\end{equation}
which has $\theta_{E_6} = \pi - \tan^{-1} \sqrt{5/3} \sim 0.71 \pi$. 
There are also the inert model with
$Q'_I = -Q'(\theta_{E_6} = \tan^{-1} \sqrt{3/5} \sim 0.21 \pi)$,
the neutral $N$ model with 
$\theta_{E_6} = \tan^{-1} \sqrt{15} \sim 0.42 \pi$, and
the secluded sector model with 
$\theta_{E_6} = \tan^{-1} \sqrt{15}/9 \sim 0.13 \pi$.
The chiral charges for each member of the $\mathbf{27}$
are also listed in the last four columns in Table~\ref{E6} for 
these four variations of $Z'$ models within $E_6$.
Here we take the assumption that all the exotic particles, other than the 
particle contents of the MSSM, are very heavy and well beyond the reaches
of all current and planned colliders. 

\begin{table}[tbh!]
\caption{\small \label{E6}
The chiral charges of the left-handed fermions for various $Z'$ bosons 
arised in $E_6$ \cite{paul}. 
Note that 
$Q'_{f_R} = - Q'(f^c)$ since all the right-handed SM fermions are necessarily
converted into left-handed charge-conjugated fermions in order to put them into
the irreducible representation of $\mathbf{27}$ of $E_6$.
}
\smallskip
\begin{ruledtabular}
\begin{tabular}{cccccccc}
$SO(16)$ & $SU(5)$ & $2\sqrt{10} Q'_\chi$  & $2\sqrt{6}Q'_\psi$ & 
$2\sqrt{15}Q'_\eta$ & $2Q'_I$ & $2\sqrt{10}Q'_N$  & $2\sqrt{15}Q'_{\rm sec}$ \\ 
\hline
$\mathbf{16} $ & $\mathbf{10}(u,d, u^c, e^c)$ & $-1$ & $1$ & $-2$ & $0$ &
                  $1$ & $-1/2$ \\
                & $\mathbf{5^*}(d^c,\nu, e^-)$ & $3$ & $1$ & $1$ & $-1$ &
                  $2$ & $4$ \\
                & $\nu^c$ & $-5$ & $1$ & $-5$ & $1$ & $0$ & $-5$ \\
\hline
$\mathbf{10} $ & $\mathbf{5}(D, H_u)$ & $2$ & $-2$ & $4$ & $0$ &
                  $-2$ & $1$ \\
                & $\mathbf{5^*}(D^c, H_d)$ & $-2$ & $-2$ & $1$ & $1$ &
                  $-3$ & $-7/2$ \\
\hline
$\mathbf{1} $ & $\mathbf{1} S $ & $0$ & $4$ & $-5$ & $-1$ &   $5$ & $5/2$ 
\end{tabular}
\end{ruledtabular}
\end{table}

\subsubsection{$B-L$ models}

The extra $U(1)$ symmetry here is the $U(1)_{B-L}$ with $Q'_{f} = \frac{1}{2}
(B-L)^f$, where $B$ and $L$ are the baryon and lepton numbers, respectively.
The theory is vector-like, thus
the chiral charges for the left- and right-handed fermions are therefore
the same, $Q'_{f_L} = Q'_{f_R} =  \frac{1}{2}(B-L)^f$
as they are summarized in Table~\ref{b-l}.

\begin{table}[h!]
\caption{\small \label{b-l}
The chiral charges of the left-handed fermions for the $Z'_{B-L}$ boson.
Note that $Q'_{f_R} = - Q'(f^c)$ since charge-conjugated fields are used here.
}
\smallskip
\begin{ruledtabular}
\begin{tabular}{cc}
             & $Q'_{B-L} = (B-L)/2$ \\
\hline
$Q = (u,d)$  &  $1/6$ \\
$u^c$        &  $-1/6$ \\
$d^c$        &  $-1/6$ \\
$L = (\nu, e^-)$ & $-1/2$ \\
$e^c$            & $1/2$ 
\end{tabular}
\end{ruledtabular}
\end{table}

\section{Decays of $Z'$}

In general, the $Z'$ boson can decay into all SM fermion, squark, slepton,
sneutrino, neutralino, chargino, and Higgs-boson pairs.  The decays into
neutralino and chargino pairs go via the couplings to Higgsinos. The 
Higgs-boson pairs include $Ah^0,\, AH^0,$ and $H^+ H^-$.  All decay modes are
subjected to kinematic threshold.  We have chosen typical input
parameters for the supersymmetric particles:
\footnote{There are additional $D$-term contributions to the sfermion
and Higgs boson masses from the breaking of the $U(1)'$ symmetry \cite{tony},
which further depend on the details of
the VEV of additional Higgs fields that break the $U(1)'$ symmetry.
Instead,  we used the physical masses as inputs to our study.}
\footnote{
We choose the squark mass to be heavy enough such that the decay into squark
pair is not open.  We have set the soft parameters $M_{\tilde{L}}^2 = 
M_{\tilde{E}^c}^2 = (200 \;{\rm GeV})^2$ such that the physical slepton
masses are about $205$ and $204$ GeV for the left-handed and right-handed
sleptons, but about 190 GeV for the sneutrino. We have ignored the
additional $D$-term contribution from the $U(1)'$.}
\begin{eqnarray}
{\rm Set (A):} \;\;\;\;
&&\tan\beta = 5\,, \;\; m_{\tilde q} = O({\rm a \; few}\;{\rm TeV})\,, \;\;
 M_{\tilde L} =  200 \;{\rm GeV}\,, \;\; 
 M_A = 500 \; {\rm GeV} \,,  \nonumber \\
&&
 M_1 = 100 \;{\rm GeV} \,, \;\;
 M_2 = 200 \;{\rm GeV} \,,  \;\; 
 \mu = 150 \;{\rm GeV} \,, \;\; A = 100\; {\rm GeV}\,. \label{susy}
\end{eqnarray}
With this set of choices the $Z'$ boson will not decay into squark pairs,
but can decay into all other sfermion pairs, neutralino and chargino pairs,
and Higgs-boson pairs.

\begin{figure}[th!]
\includegraphics[angle=270,width=3.2in]{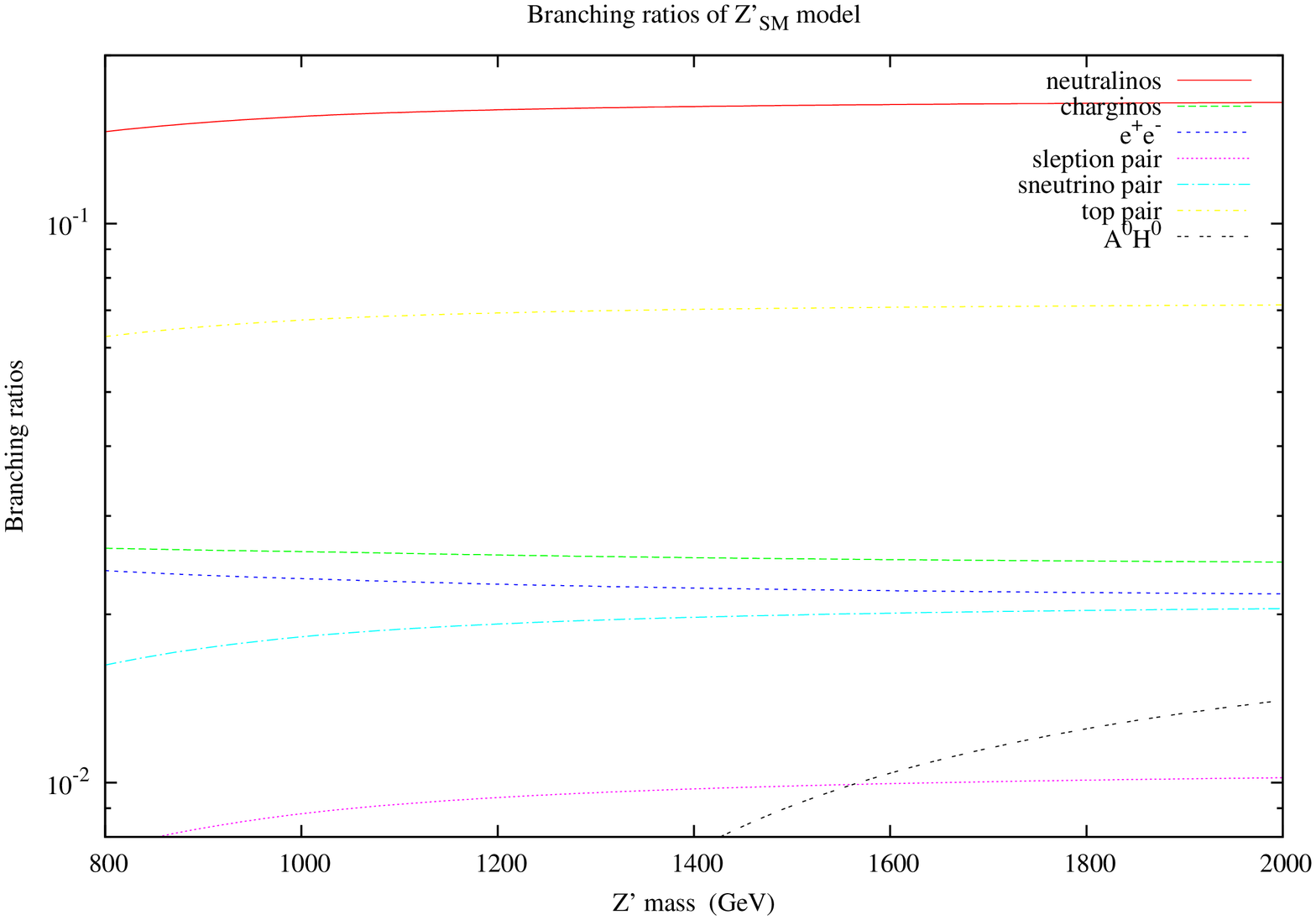}
\includegraphics[angle=270,width=3.2in]{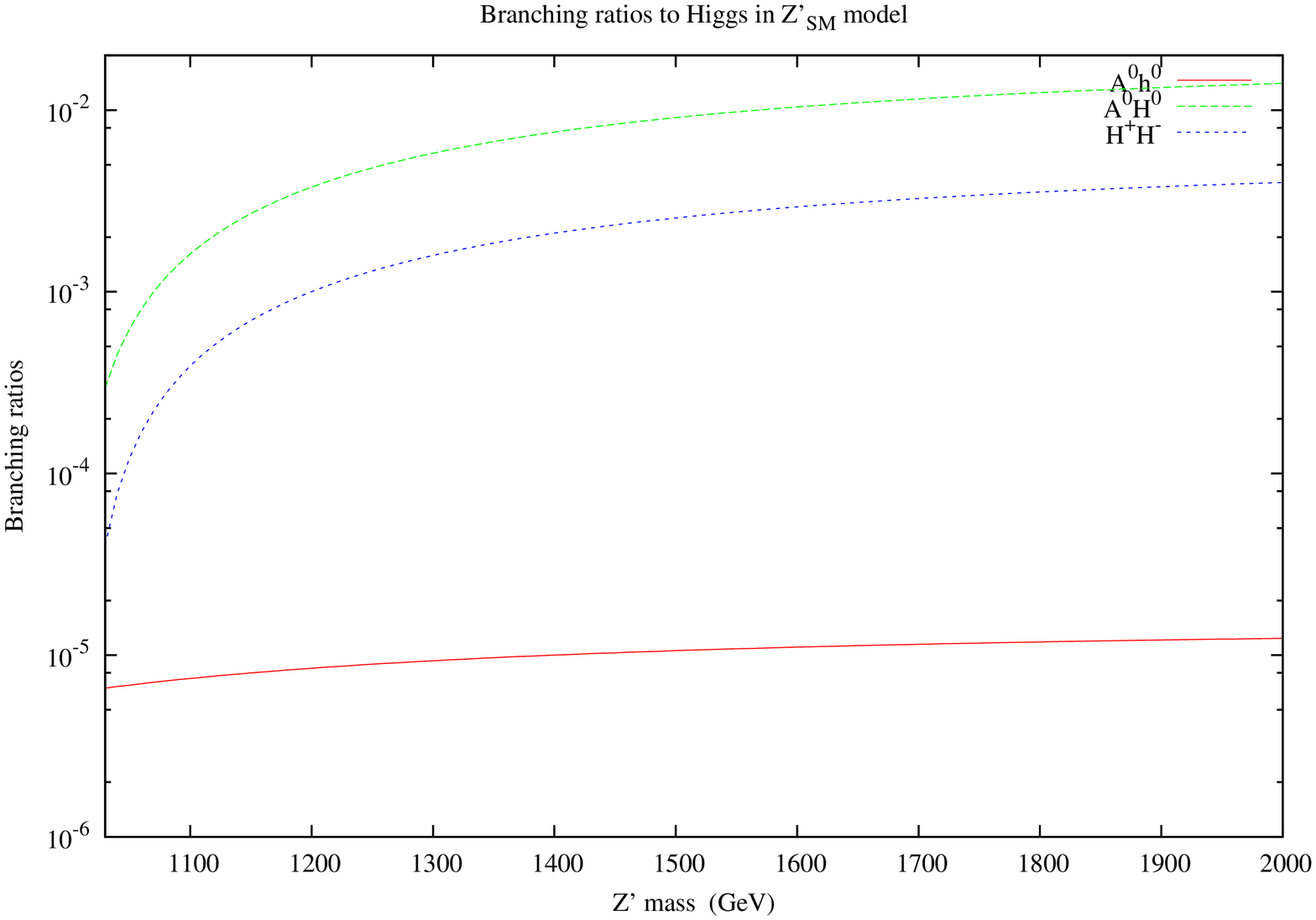}
\caption{\small \label{z-sm}
Decay branching ratios for some decay modes of the sequential $Z'$ model.}
\end{figure}

\begin{figure}[th!]
\includegraphics[angle=270,width=3.2in]{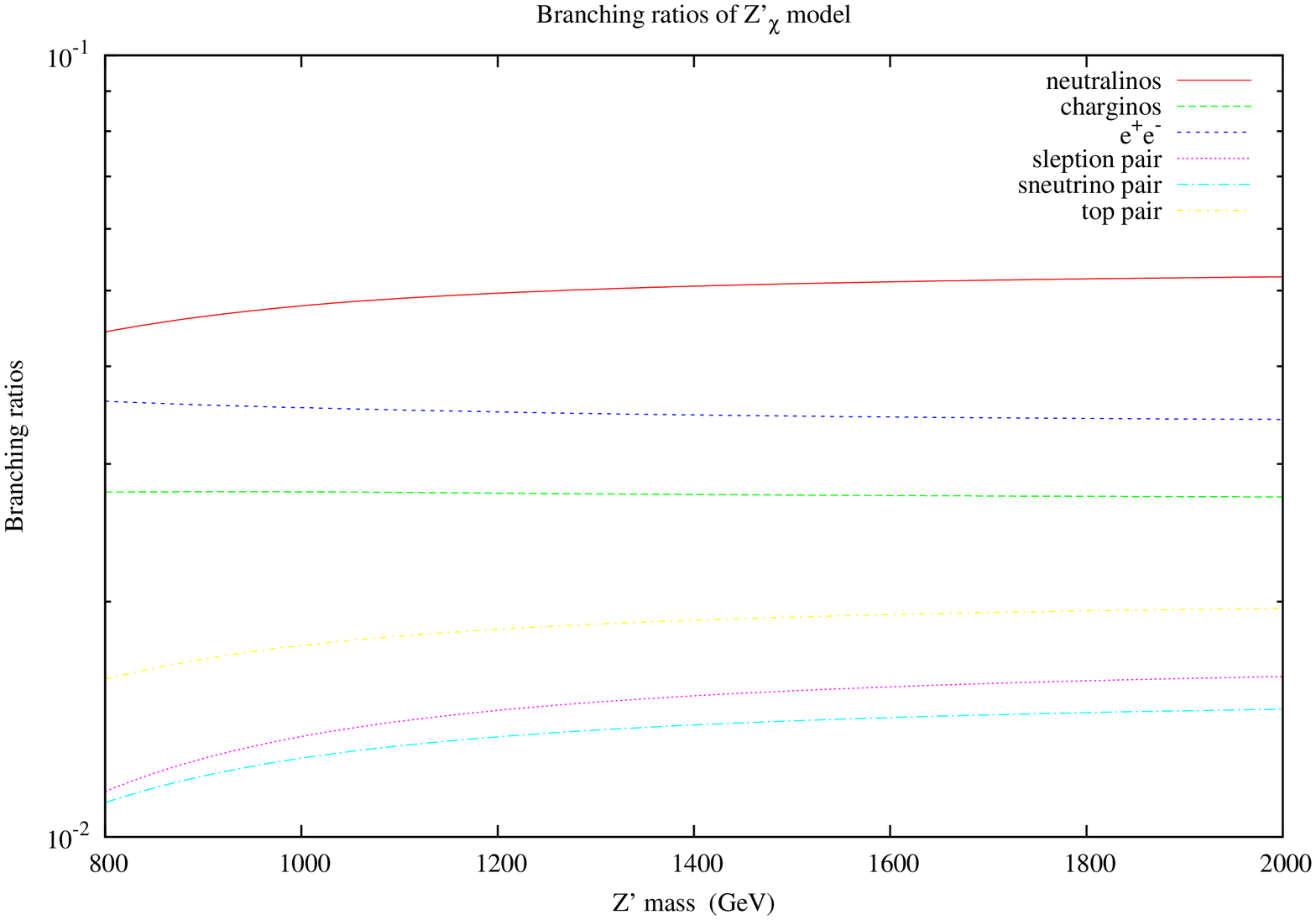}
\includegraphics[angle=270,width=3.2in]{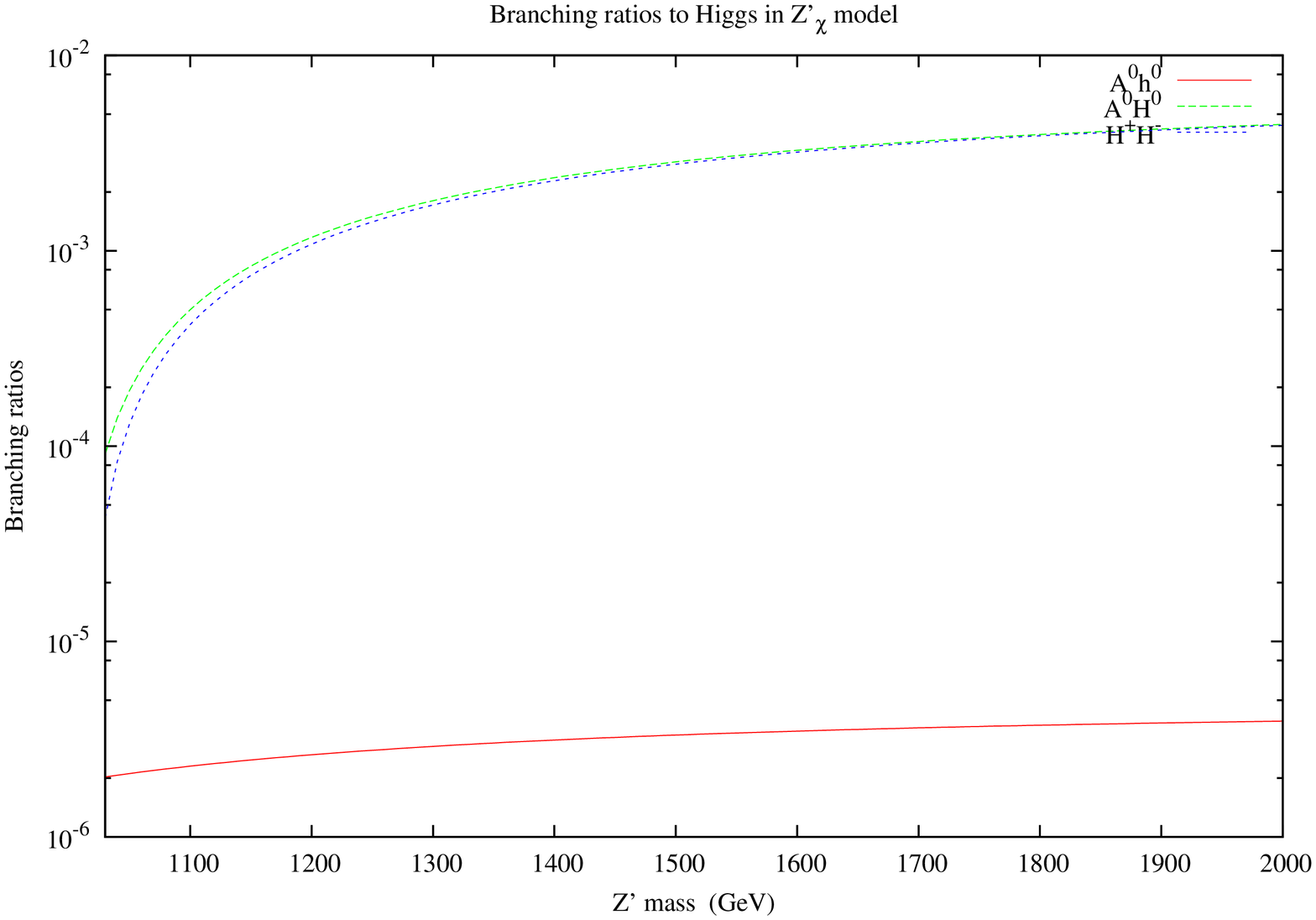}
\caption{\small \label{z-chi}
Same as Fig.~\ref{z-sm} but for $Z'_\chi$ model.}
\end{figure}

\begin{figure}[th!]
\includegraphics[angle=270,width=3.2in]{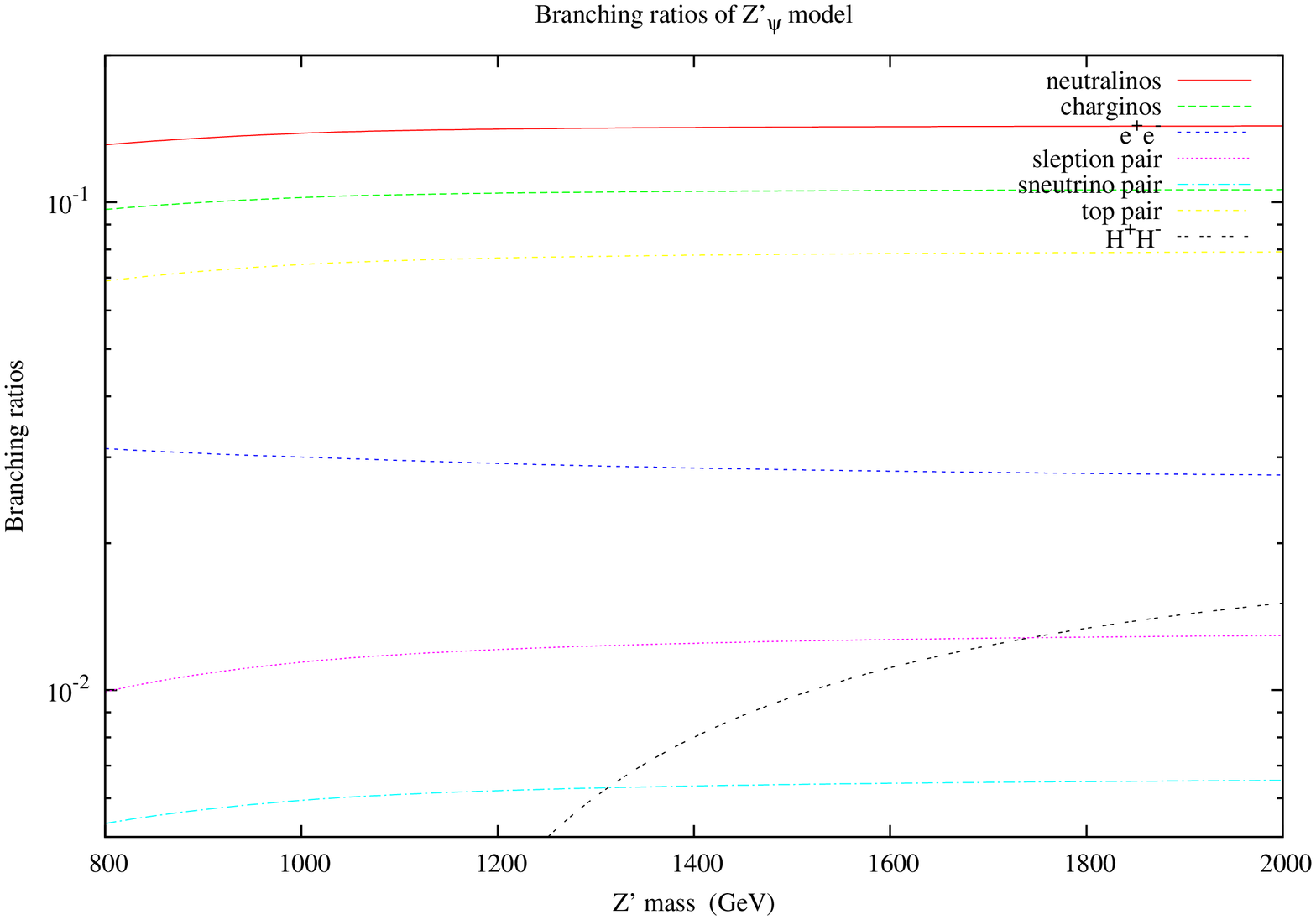}
\includegraphics[angle=270,width=3.2in]{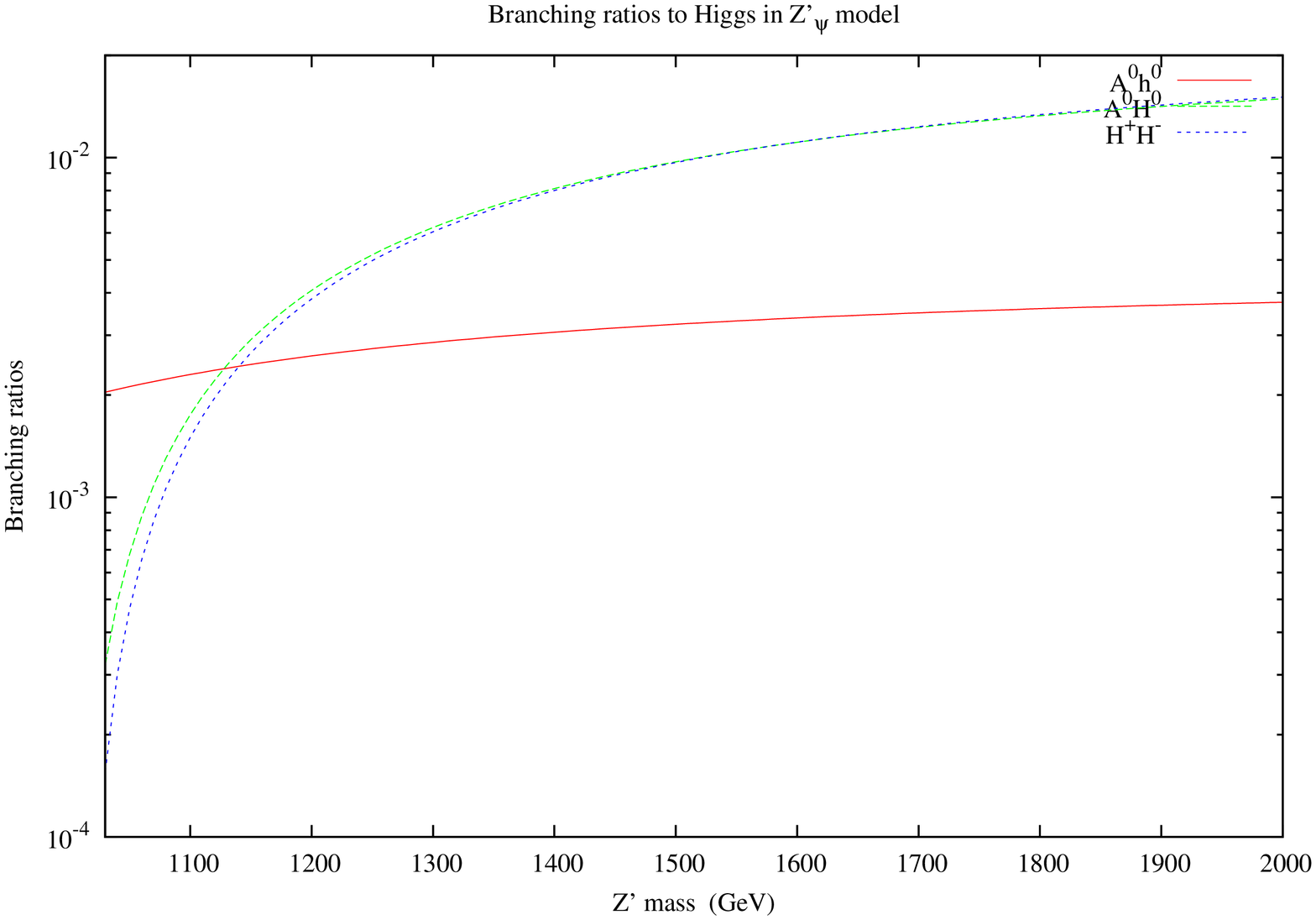}
\caption{\small \label{z-psi}
Same as Fig.~\ref{z-sm} but for $Z'_\psi$ model.}
\end{figure}

\begin{figure}[th!]
\includegraphics[angle=270,width=3.2in]{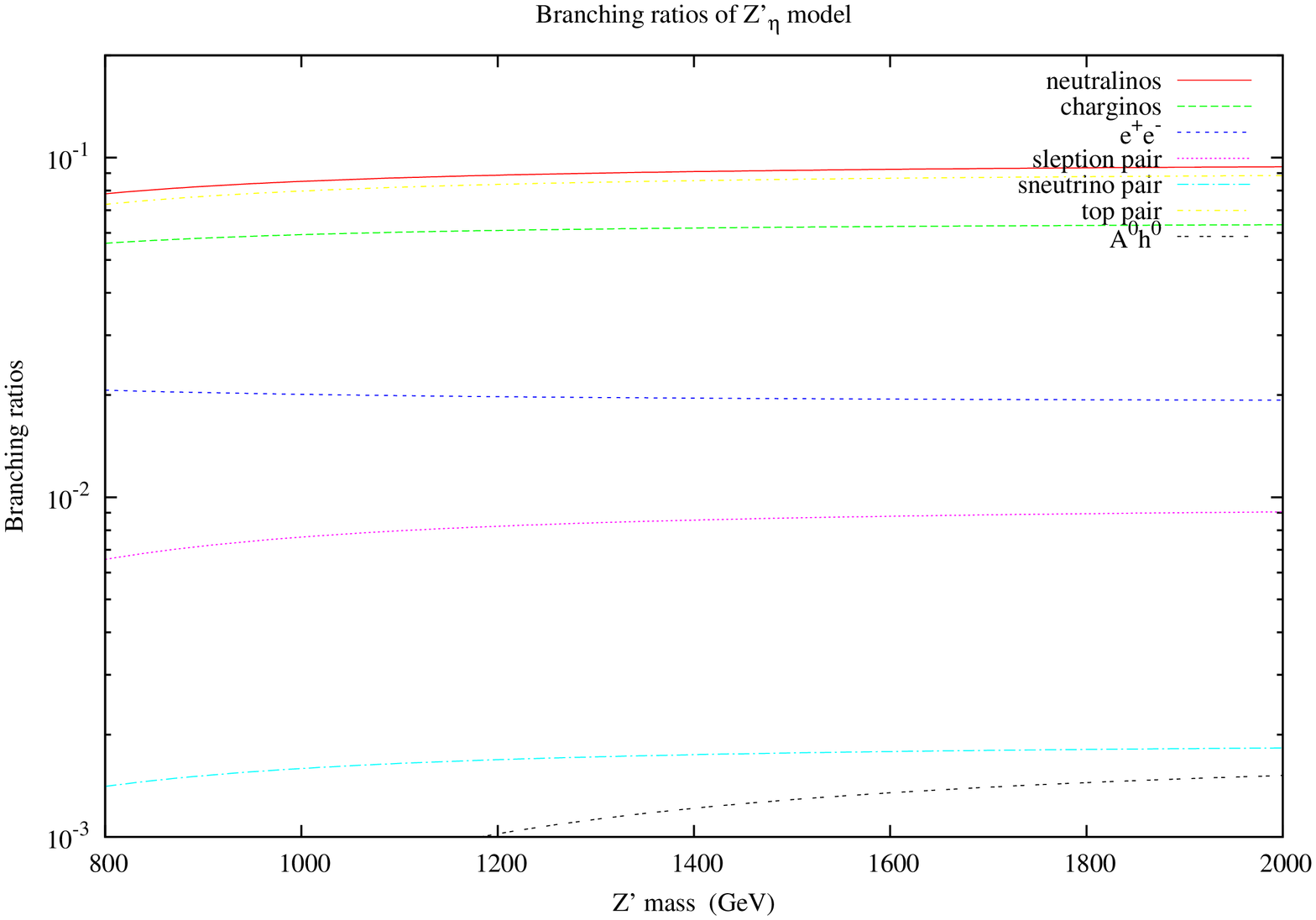}
\includegraphics[angle=270,width=3.2in]{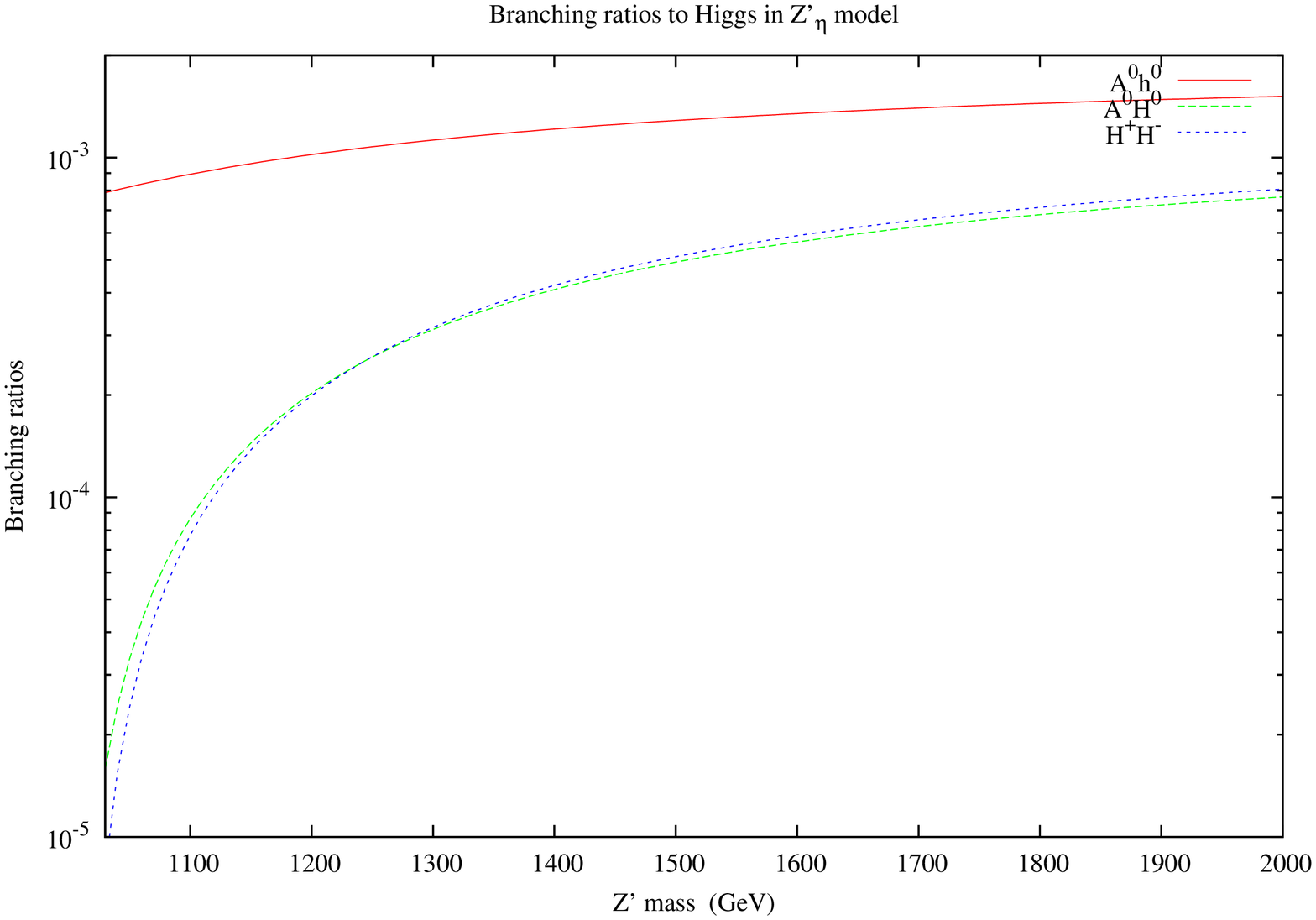}
\caption{\small \label{z-eta}
Same as Fig.~\ref{z-sm} but for $Z'_\eta$ model.}
\end{figure}

\begin{figure}[th!]
\includegraphics[angle=270,width=3.2in]{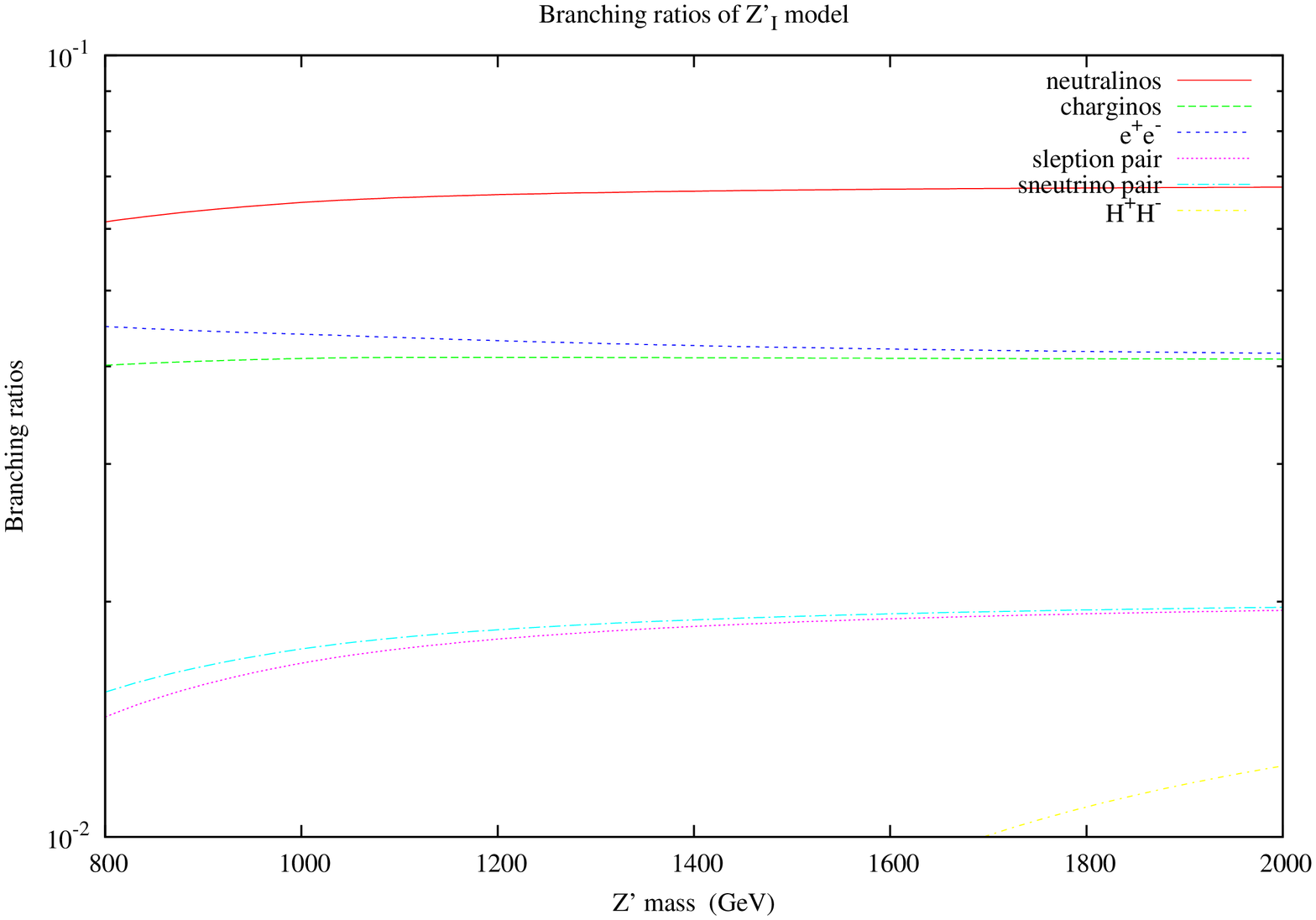}
\includegraphics[angle=270,width=3.2in]{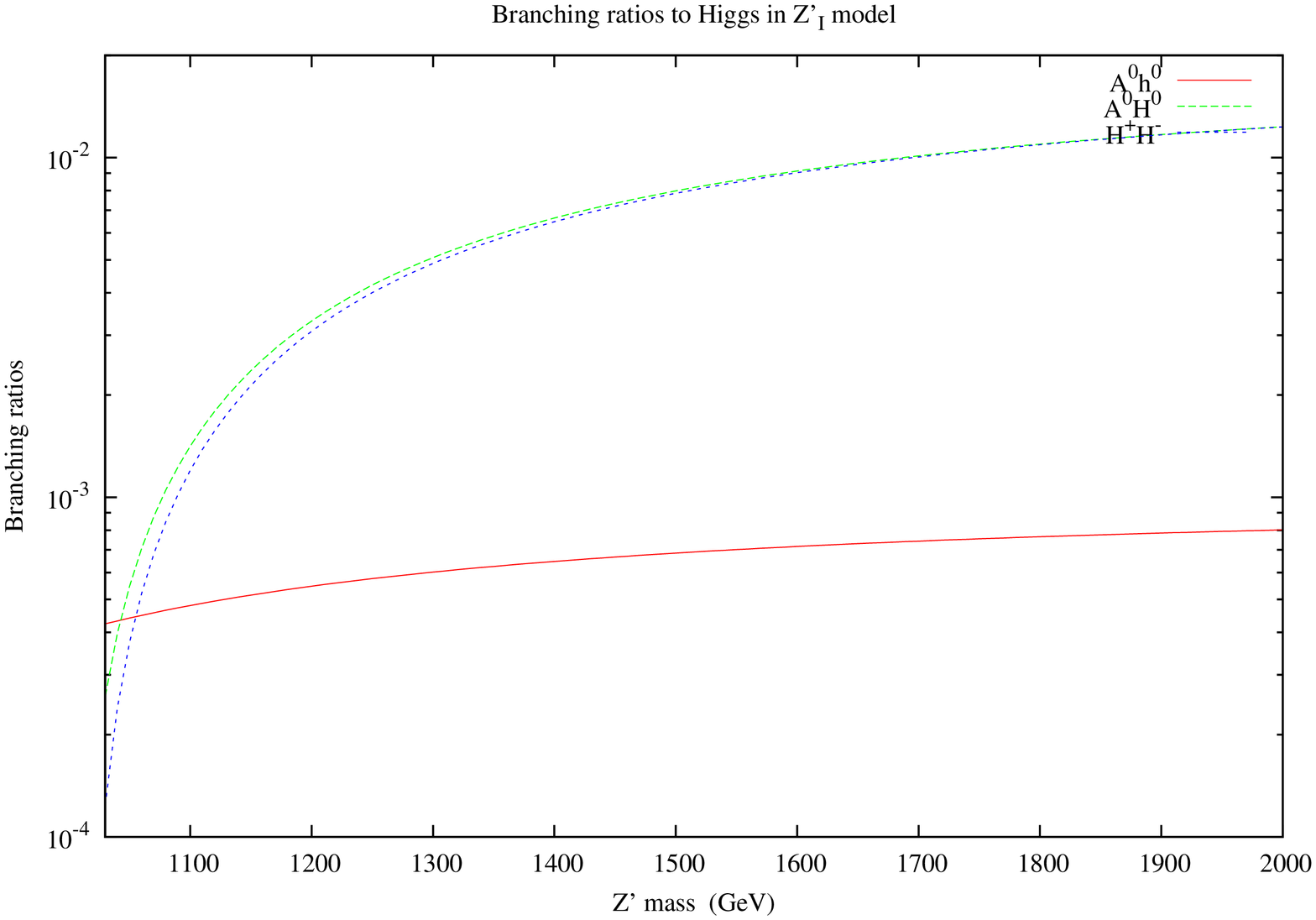}
\caption{\small \label{z-iner}
Same as Fig.~\ref{z-sm} but for $Z'_{I}$ model.}
\end{figure}

\begin{figure}[th!]
\includegraphics[angle=270,width=3.2in]{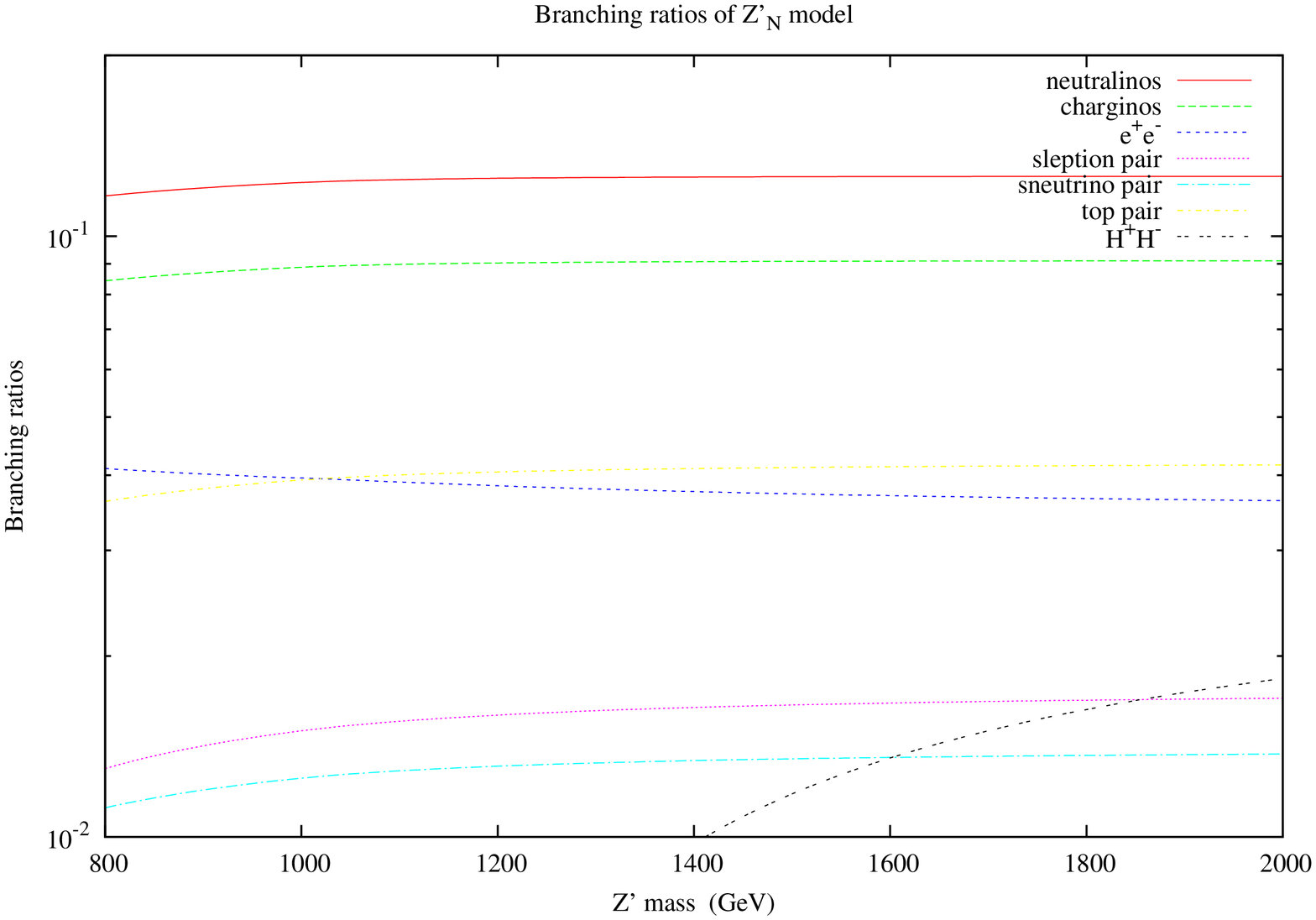}
\includegraphics[angle=270,width=3.2in]{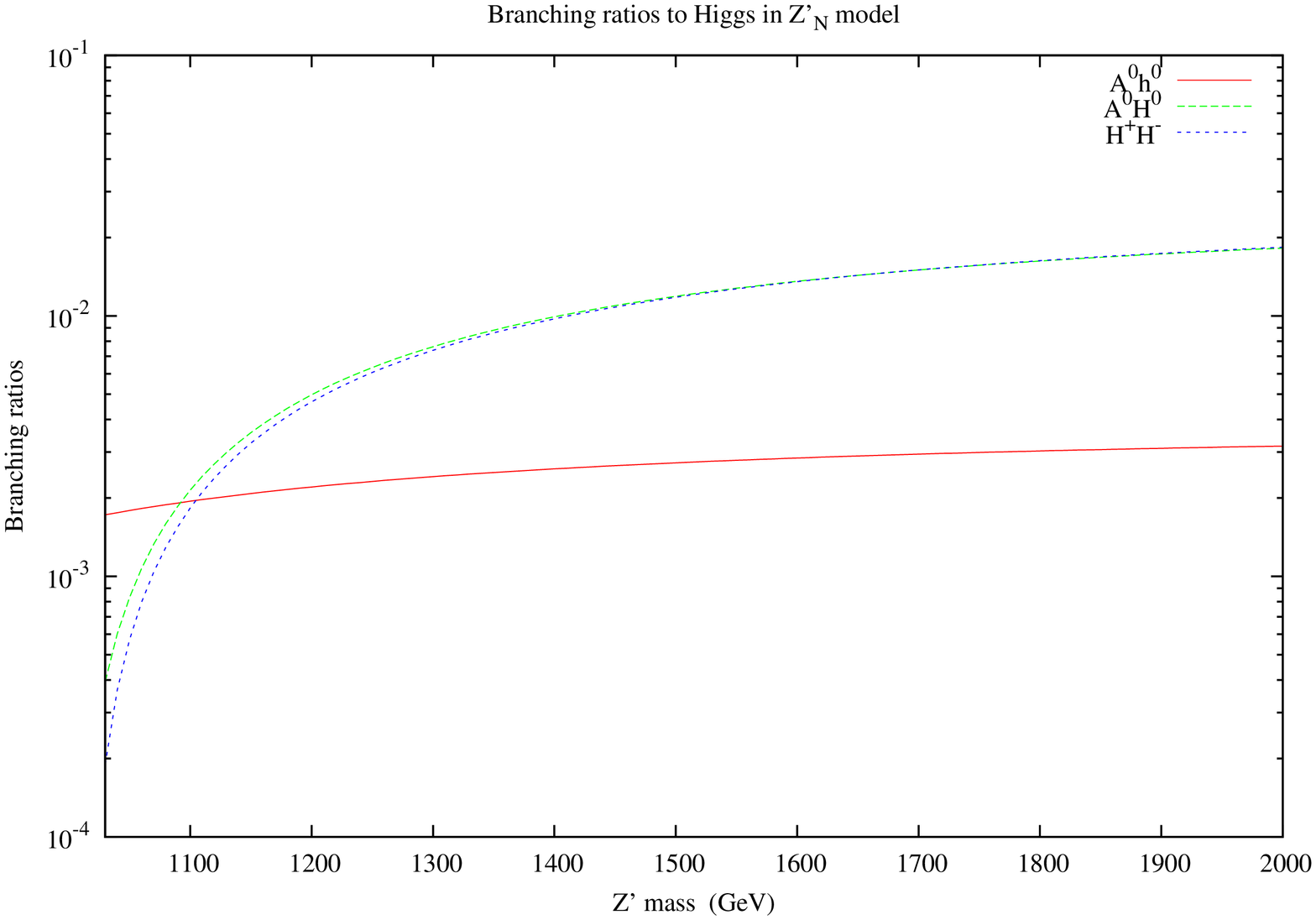}
\caption{\small \label{z-N}
Same as Fig.~\ref{z-sm} but for $Z'_N$ model.}
\end{figure}

\begin{figure}[th!]
\includegraphics[angle=270,width=3.2in]{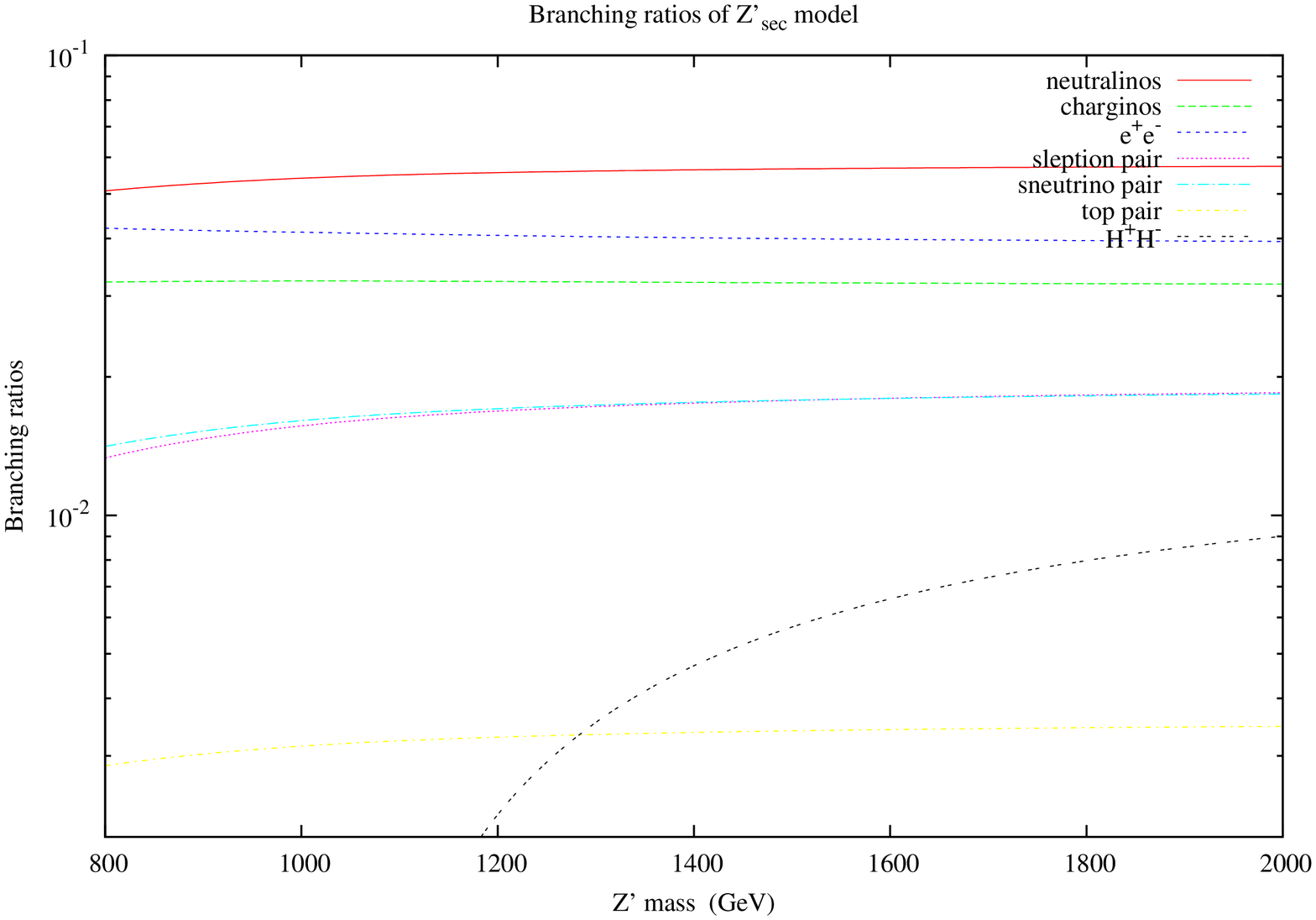}
\includegraphics[angle=270,width=3.2in]{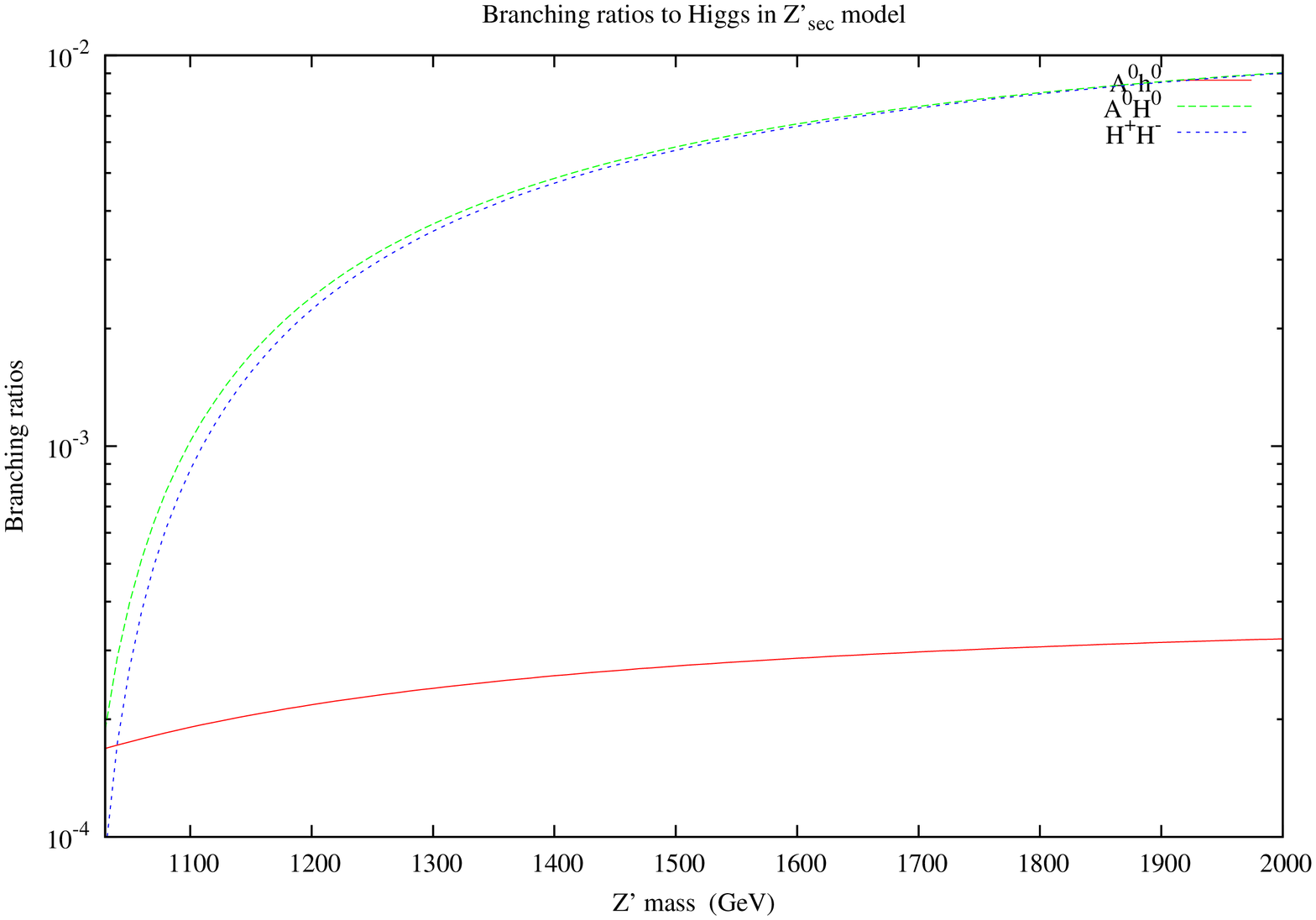}
\caption{\small \label{z-sec}
Same as Fig.~\ref{z-sm} but for $Z'_{\rm sec}$ model.}
\end{figure}

\begin{figure}[th!]
\includegraphics[angle=270,width=3.2in]{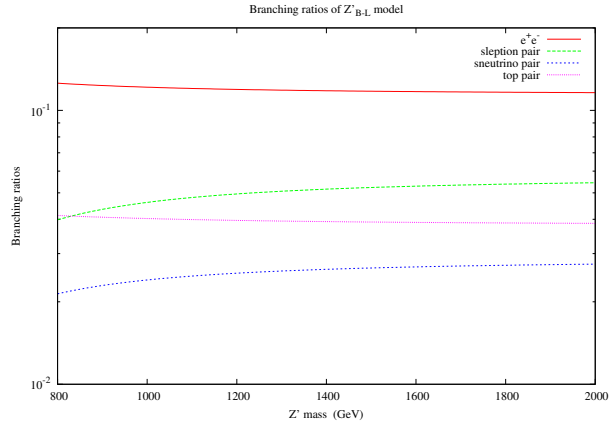}
\caption{\small \label{z-bl}
Same as Fig.~\ref{z-sm} but for $Z'_{B-L}$ model.}
\end{figure}

We show in Figs.~\ref{z-sm} to \ref{z-bl} 
the decay branching ratios of various $Z'$ bosons.
In these figures, we only 
show $e^+ e^-$ -- one of the charged lepton modes, which is the most
direct discovery mode of the $Z'$ boson, 
$\tilde{e}_L \tilde{e}_L^* + \tilde{e}_R \tilde{e}_R^* $ -- one of the
slepton modes, 
$\tilde{\nu}_{eL} \tilde{\nu}_{eL}^* $ -- one of the sneutrino modes,
but the sum of neutralino and chargino pairs. The other charged-lepton 
modes and slepton modes are approximately the same as the corresponding
one shown in the figures. The top-quark pair and the other Higgs-boson 
pairs are also shown.  
Note that the $Z'_{B-L}$ does not couple to the Higgs fields.

We also select another set of SUSY parameters that we call set (B), typically
it is a large $\tan\beta$ case.
\begin{eqnarray}
{\rm Set (B):} \;\;\;\;
&&\tan\beta = 40\,, \;\; m_{\tilde q} = O({\rm a \; few}\;{\rm TeV})\,, \;\;
 M_{\tilde L} =  300 \;{\rm GeV}\,, \;\; 
 M_A = 500 \; {\rm GeV} \,,  \nonumber \\
&&
 M_1 = 200 \;{\rm GeV} \,, \;\;
 M_2 = 400 \;{\rm GeV} \,,  \;\;
 \mu = -300 \;{\rm GeV}\,, \;\;A = 300 {\rm GeV}\,. \label{susyB}
\end{eqnarray}
We shall show the results for the limits from the 
Tevatron and the sensitivities at the LHC with the two choices of set (A) and set (B).

\section{Current Limits from the Tevatron}

\begin{figure}[t!]
\includegraphics[angle=270,width=3.2in]{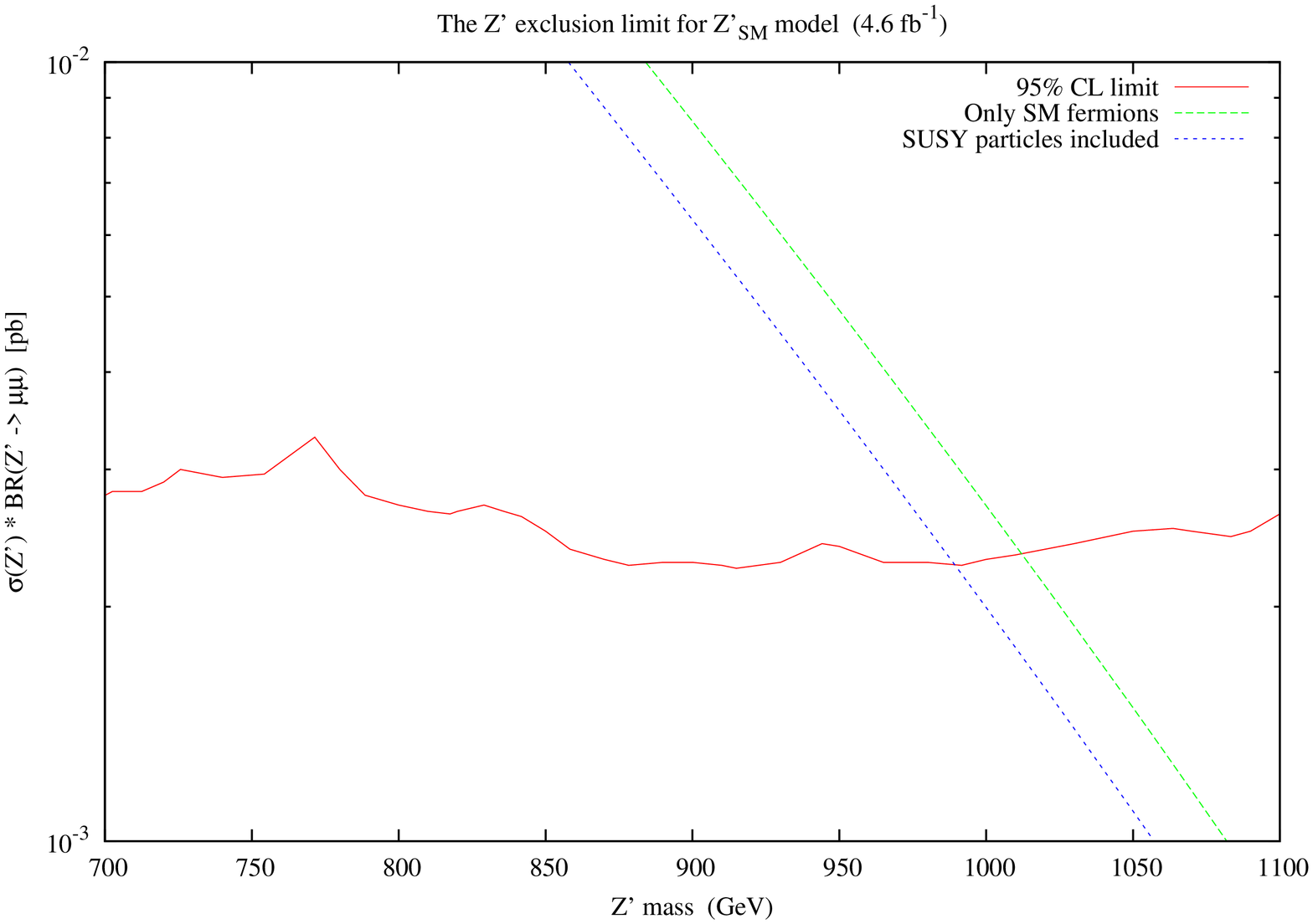}
\includegraphics[angle=270,width=3.2in]{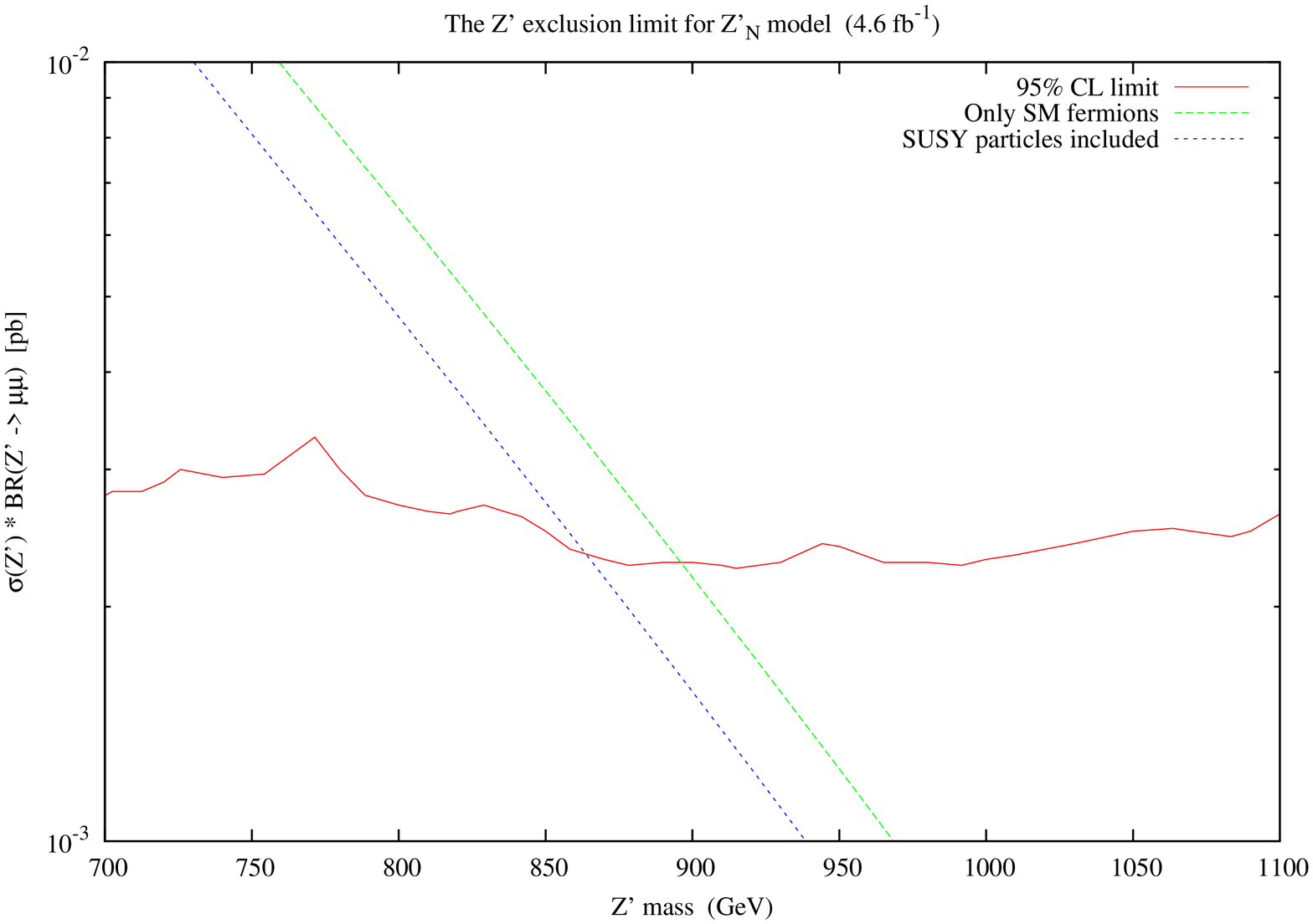}
\caption{\small \label{z_sm} 
Exclusion limits on (a) the sequential SM $Z'$ boson and (b)
the $Z'_N$ boson.  The lines are
the theoretical production cross section of $\sigma(Z') \times
B(Z' \to \mu^+ \mu^-)$ for (i) decays into SM fermions only and 
(ii) decays into SM fermions and SUSY particles. The solid (red) line
shows the most current upper limit from the CDF.
}
\end{figure}

\begin{figure}[thb!]
\includegraphics[angle=270,width=5in]{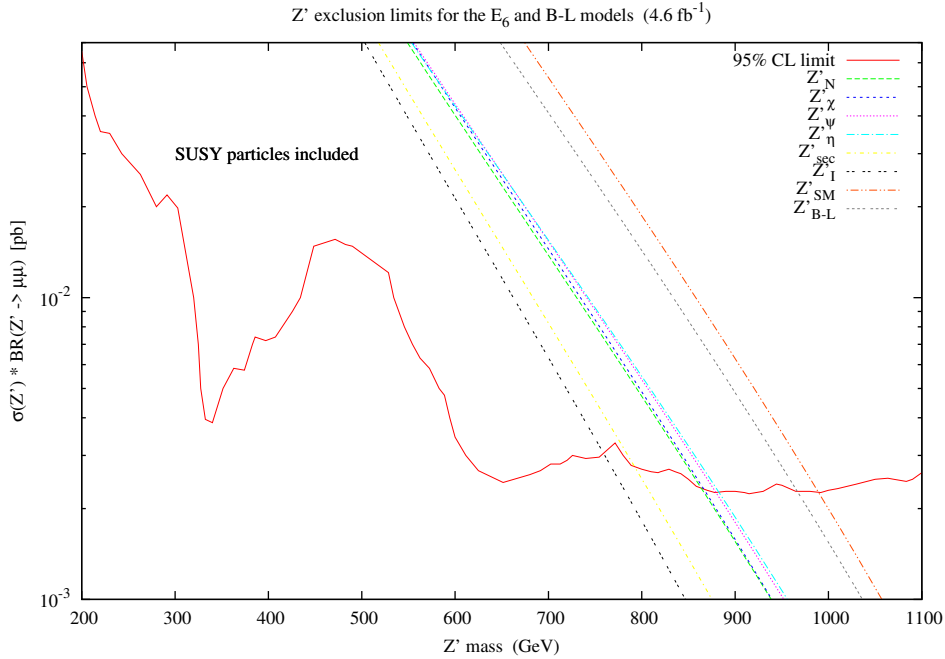}
\caption{\small \label{fig-susy} 
Exclusion limits on $Z'$ bosons of various models described in
Sec. II when the SUSY decay modes are open. The SUSY parameters chosen
are listed in Eq.~(\ref{susy}).
The lines are the theoretical production cross section 
of $\sigma(Z') \times B(Z' \to \mu^+ \mu^-)$.  The solid (red) line
shows the most current upper limit from the CDF.
}
\end{figure}

The current limits quoted from the Tevatron \cite{cdf,d0} and 
the LHC \cite{atlas} are based on the theoretical calculation of
$\sigma(Z') \times B(Z' \to \ell^+ \ell^-)$, where $\ell = e$ or $\mu$.
There the decay of $Z'$ includes only the SM fermions.
The most updated limits from the Tevatron range from about $0.8$ 
to about $1$ TeV, depending on the models. 
Here we are going to demonstrate that because of the additional decay
channels of the $Z'$, including the sfermions, neutralinos, and charginos,
the mass limits from the Tevatron are reduced by a nontrivial amount.
From the plots shown in the last section we saw that 
the branching ratio into charged
leptons decreases substantially because of the opening of the SUSY modes.
Therefore, the product $\sigma(Z') \times B(Z' \to \ell^+ \ell^-)$ decreases.
We show in Fig.~\ref{z_sm} the limits for (a) the sequential $Z'$ boson
and (b) the $Z'_N$ boson
for decays into SM fermions only and for decays into SM fermions and
SUSY particles. The choice of SUSY parameters is given in 
Eq.~(\ref{susy}) for set (A).
With these choices the $Z'$ cannot decay into squarks but can decay into
slepton pairs, sneutrino pairs, neutralino and chargino pairs, plus
the original SM fermions.  Note that the $Z'$ can also decay into
Higgs-boson pairs such as $A h^0,\; A H^0$, and $H^+ H^-$.  The limit 
for the sequential $Z'$ boson shifts from
1012 GeV down to 989 GeV (a difference of 23 GeV).  The other $Z'$ models
behave similarly.  
We show the $Z'$ mass limits for various
models when the SUSY decay modes are open in 
Fig.~\ref{fig-susy}.
We summarize the changes in limits in 
Table~\ref{table-limit} for the parameter sets (A) and (B).
Compared with the case of SM fermions only, the reduction in the 
$Z'$ mass limits ranges between
11 (5) and 32 (18) GeV for various models with the SUSY parameters of set (A)
(set (B)).

\begin{table}[thb!]
\caption{\small \label{table-limit}
Table showing the down shift of the limits on the $Z'$ boson mass due
to the inclusion of the SUSY decay modes of $Z'$.
The choices of SUSY parameters are given in set (A) and in set (B).
}
\smallskip
\begin{ruledtabular}
\begin{tabular}{cccccccccc}
& $Z'_N$ & $Z'_\chi$ & $Z'_\psi$ & $Z'_\eta$ & $Z'_{\rm sec}$ & $Z'_I$ &
  $Z'_{\rm SM}$ & $Z'_{B-L}$ \\
\hline
SM fermions only & 896 & 882 & 910 & 898 & 811 & 770 & 1012 & 987  \\
SUSY included [Set (A)]  & 864 & 866 & 879 & 882 & 793 & 759 &  989 & 966  \\
SUSY included [Set (B)]  & 880 & 874 & 892 & 890 & 802 & 765 &  997 & 973
\end{tabular}
\end{ruledtabular}
\end{table}

\section{Projected sensitivities at the LHC}

Here we investigate the sensitivities at the LHC with $\sqrt{s} = 7, 10$, and 14 TeV with projected luminosities of $1,\,10$, and $100$ fb$^{-1}$, respectively,
using the conventional search channel, $pp \to Z' \to \ell^+ \ell^-$.
Since the $Z'$ boson is a narrow resonance state, we adopt the following
procedures to obtain the $5\sigma$ discovery potential of each $Z'$ model.

\begin{enumerate}
\item 
 Take a $Z'$ boson of mass, say, 800 GeV.  We pick a conservative bin-size 
resolution of order 100 GeV in the $M_{\ell\ell}$ distribution around TeV.
Therefore, we are looking at the window $(800 - 50, 800+ 50)$ GeV.

\item We calculate the expected Drell-Yan background in this mass window
with the center-of-mass energy and the corresponding luminosity, say $N$ 
events.  

\item If $N > 10$ we can use Gaussian distribution, and the standard
deviation is given by $\sqrt{N}$.  The $5\sigma$ signal corresponds to
$5 \times \sqrt{N}$ events.   If $N<10$ then we should use Poisson
statistics to determine the $5\sigma$ signal events.

\item  Check the $\sigma(Z') \times B(Z' \to \ell^+ \ell^-)$ if it is 
larger or smaller than the $5\sigma$ events.  If 
 $\sigma(Z') \times B(Z' \to \ell^+ \ell^-)$ is larger, it means that
the discovery sensitivity can go further;  otherwise, 
the discovery sensitivity cannot go further
and the current $M_{Z'}$ is the $5\sigma$ discovery limit, and the
procedures stop here.

\item Repeat with an increment of, say 10 GeV, for $M_{Z'}$ 
and go back to step (1).
\end{enumerate}

The $5\sigma$ discovery sensitivity for LHC-7, LHC-10, and LHC-14 with
luminosities $1,\,10$, and $100$ fb$^{-1}$, respectively, are shown in
Table~\ref{lhc}
for the decay into SM particles only, and for decay into
SM and SUSY particles with SUSY parameter sets (A) and (B).
The difference in sensitivity reach between set (A) and (B) is very small here,
because the mass of the $Z'$ boson near the sensitivity reach is so large 
that the masses of the sleptons and gauginos involved are relatively negligible.
The mass reach for the $E_6$ $Z'$ bosons is about
$1.3-1.5$ TeV at the LHC-7 with 1 fb$^{-1}$, and could be up to $2.5 -
2.6$ TeV at the LHC-10 with 10 fb$^{-1}$, and $4.2 - 4.3$ TeV at the
LHC-14 with 100 fb$^{-1}$ 
when supersymmetry is included in the decay of the $Z'$.
On the other hand, the mass reach for
$Z'_{\rm SM}$ and $Z'_{B-L}$ is about one to a few hundred GeVs better
than the $Z'$ in $E_6$ models.  
We have checked that the number of signal events at the mass-reach
limit, say for the $Z'_N$ model, is 5 at the LHC-7 against a Drell-Yan
background of $0.09$ events, and is 4 at the LHC-10 and LHC-14 against
a Drell-Yan background of $0.04$ and $0.03$ events, respectively.  The
probability for background fluctuation is $\alt 10^{-7}$, such that it
is a $5\sigma$ discovery.  Note that we only count one channel of the
charged lepton (electron).  If both electron and muon are counted,
there will be $8-10$ charged-lepton pair events for discovery, which
should be clean enough against negligible Drell-Yan background.

\begin{table}[tbh!]
\caption{\small \label{lhc}
The $5\sigma$ discovery sensitivity reach in $M_{Z'}$ (TeV) 
for various $Z'$ models
at the LHC-7, -10, and -14
with luminosities $1,\,10$, and $100$ fb$^{-1}$, respectively.
The cases with decay into SM particles only, and decay into SM and 
SUSY particles with parameter sets (A) and (B) are shown.
}
\medskip
\begin{ruledtabular}
\begin{tabular}{cccccccccc}
 & $Z'_N$ & $Z'_\chi$ & $Z'_\psi$ & $Z'_\eta$ & $Z'_{\rm sec}$ & 
      $Z'_I$ &  $Z'_{\rm SM}$ & $Z'_{B-L}$ \\
\hline
      & \multicolumn{8}{c}{ LHC-7 at 1 fb$^{-1}$ } \\
SM              & 1.50 & 1.53 & 1.46 & 1.35  & 1.53  & 1.52 & 1.83 & 1.74 \\
SUSY [Set (A)]  & 1.36 & 1.49 & 1.32 & 1.31  & 1.48  & 1.46 & 1.68 & 1.60 \\
SUSY [Set (B)]  & 1.37 & 1.49 & 1.33 & 1.31  & 1.48  & 1.47 & 1.69 & 1.61 \\
\hline
      & \multicolumn{8}{c}{ LHC-10 at 10 fb$^{-1}$ } \\
SM             & 2.65 & 2.69 & 2.61 & 2.53  & 2.67  & 2.65 & 3.16 & 3.03 \\
SUSY [Set (A)] & 2.50 & 2.62 & 2.45 & 2.45  & 2.59  & 2.56 & 3.03 & 2.82 \\
SUSY [Set (B)] & 2.51 & 2.63 & 2.46 & 2.46  & 2.59  & 2.56 & 3.04 & 2.83 \\
\hline
      & \multicolumn{8}{c}{ LHC-14 at 100 fb$^{-1}$ } \\
SM             & 4.47 & 4.51 & 4.44 & 4.34  & 4.44  & 4.41 & 5.21 & 5.04 \\
SUSY [Set (A)] & 4.25 & 4.41 & 4.21 & 4.23  & 4.33  & 4.27 & 5.03 & 4.89 \\
SUSY [Set (B)] & 4.26 & 4.41 & 4.22 & 4.23  & 4.33  & 4.27 & 5.03 & 4.89
\end{tabular}
\end{ruledtabular}
\end{table}

\section{Supersymmetric Decay Modes}

Identification of supersymmetric decay modes of the $Z'$ has it own
interests, namely, to understand the role of $Z'$ in the SUSY
breaking.  Moreover, if the $Z'$ boson decays frequently into
SUSY particles, we can make use of the SUSY channels to probe for the
$Z'$.  So far, in the models that we illustrate the branching ratio
into charged leptons is not negligible, such that the best discovery mode
is still the charged-lepton mode, which cleanly shows the peak in the
invariant-mass distribution.  Nevertheless, there exist models, e.g,
Refs.~\cite{stucek}, in which the charged-lepton decay mode is highly
suppressed. One could also imagine that a $Z'$ does not couple to fermions
or sfermions but only to the Higgs sector, such that it couples solely to Higgs
bosons and Higgsinos.  In such an extreme the $Z'$ would substantially
decay into Higgsinos (or the physical neutralinos and charginos after mixings).
In other words, the supersymmetric decay modes of the $Z'$ boson could be
sizable and useful for understanding the SUSY breaking. 

Typically, the SUSY decay modes include 
(i) $Z' \to \tilde{\ell} \tilde{\ell}^* \to \ell^+ \ell^- 
\tilde{\chi}^0_1 \tilde{\chi}^0_1$, 
(ii) $Z' \to \tilde{\chi}^0_1 \tilde{\chi}^0_2 \to \ell^+ \ell^- 
\tilde{\chi}^0_1 \tilde{\chi}^0_1$, 
(iii) $Z' \to \tilde{\chi}^+_1 \tilde{\chi}^-_1 \to \ell^+ \ell^- \nu \bar \nu 
\tilde{\chi}^0_1 \tilde{\chi}^0_1$,  etc.
Such leptonic modes give rise to a signature consisting of a 
charged-lepton pair and large missing energies. 
It is clean and we can construct the cluster transverse
mass $M_T$ of the lepton pair and the missing energy. The transverse mass
would indicate a broad peak structure, which is sensitive to 
the intermediate $Z'$ boson mass.  Here we only give a taste of what one
can do to see the presence of the $Z'$ via the supersymmetric decays. 
Detailed studies including various SUSY spectra and decay modes, and 
branching ratios will be given in a future publication. 

\subsection{Slepton-pair production}

\begin{figure}[h!]
\centering
\includegraphics[width=5in]{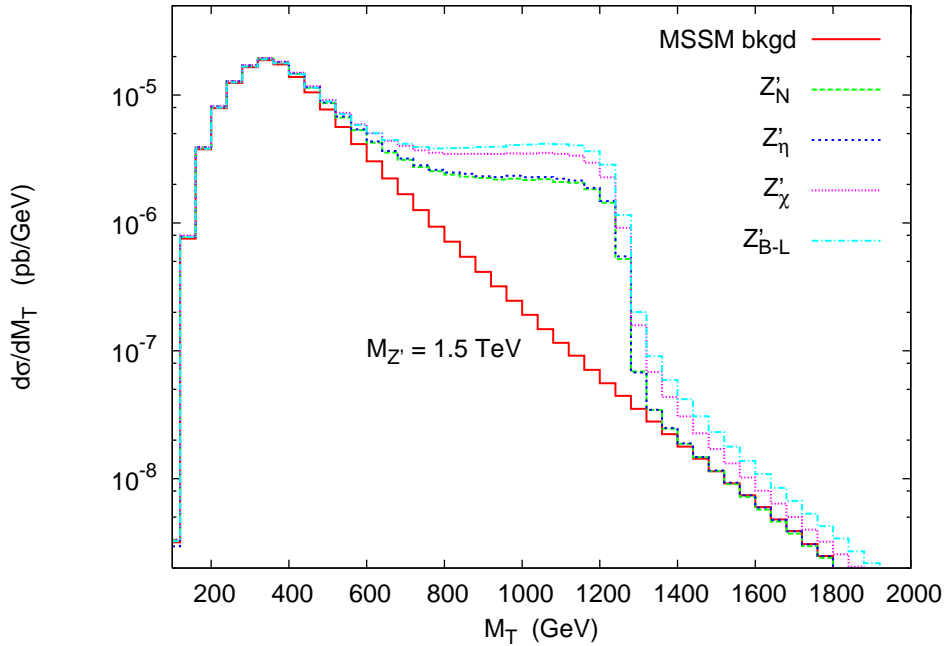}
\medskip
\caption{\small \label{mt}
The differential cross section versus the cluster transverse mass $M_T$
defined in Eq.~(\ref{cluster})
for $p p \to \gamma,Z,Z' \to \tilde{e}_L \tilde{e}_L^* +
\tilde{e}_R \tilde{e}_R^* \to
e^+ e^- \widetilde{\chi}_1^0 \widetilde{\chi}_1^0$.  
We have applied the leptonic cuts
given in Eq.~(\ref{leptoncut}).  Both $\tilde{e}_L \tilde{e}_L^*$
and $\tilde{e}_R \tilde{e}_R^*$ are included. The underneath curve
is without the $Z'$ boson while the upper curve includes the $Z'$ boson.
The $Z'$ models shown here are $Z'_N, Z'_\eta , Z'_\chi ,$ 
and $Z'_{B-L}$. The masses are 
$M_{Z'} =1.5$ TeV, $m_{\tilde{e}_{L,R}} = 200$ GeV, and $m_{\widetilde{\chi}_1^0}
= 80$ GeV.
}
\end{figure}

Let us first investigate slepton-pair production in MSSM and in MSSM plus
a $Z'$.  Electroweak production of $\tilde{e}_L \tilde{e}_L^*$
or  $\tilde{e}_R \tilde{e}_R^*$ goes through the $\gamma,\;Z,\;Z'$ 
exchanges 
\footnote{In this work, we will ignore all the Higgs exchange 
diagrams since they are suppressed by light quark masses.}.
The differential cross section for the subprocess
$ q \bar q\to \tilde{e}_\alpha \tilde{e}_\alpha^*$ ($\alpha=L,R)$ is given by
\begin{equation}
\frac{d \hat \sigma}{d \cos\hat \theta} = \frac{\beta}{96\pi \hat s}\,
\left( |M_{L\alpha}|^2 + |M_{R\alpha}|^2 \right )\, 
( \hat u \hat t - m_{\tilde{e}_\alpha}^4 )  \,,
\end{equation}
where 
\begin{eqnarray}
M_{L\alpha} &=& \frac{ e^2 Q_q Q_{\tilde{e}_\alpha}}{\hat s} + 
   \frac{g_1^2}{ \hat s - m_Z^2 + i m_Z \Gamma_Z} g_L^q g^{\tilde{e}_\alpha}
 + \frac{g_2^2}{ \hat s - m_{Z'}^2 + i m_{Z'} \Gamma_{Z'}} 
  Q'_{q_L} Q'_{\tilde{e}_\alpha} \; , \\
M_{R\alpha} &=& \frac{ e^2 Q_q Q_{\tilde{e}_\alpha}}{\hat s} + 
   \frac{g_1^2}{ \hat s - m_Z^2 + i m_Z \Gamma_Z} g_R^q g^{\tilde{e}_\alpha}
 + \frac{g_2^2}{ \hat s - m_{Z'}^2 + i m_{Z'} \Gamma_{Z'}} 
  Q'_{q_R} Q'_{\tilde{e}_\alpha} \;,
\end{eqnarray}
and $\beta = \sqrt{ 1 - 4 m_{\tilde{e}_\alpha}^2/\hat s}$. 
Here the electric charge $Q_{\tilde{e}_\alpha} = Q_e$, 
the $Z$ charge $g^{\tilde{e}_\alpha} = g_{\alpha}^e$, and
the $Z'$ charge $Q'_{\tilde{e}_\alpha} = Q'_{e_\alpha}$.
The subprocess cross section is then folded with parton distribution 
functions to obtain the total cross section.  The so-produced 
$\tilde{e}_\alpha \tilde{e}_\alpha^*$ will decay into the electron and
positron and the lightest neutralinos under the normal hierarchy of
SUSY masses. 
\footnote{Here we assume the lightest neutralino is the lightest 
supersymmetric particle (LSP) and the usual order of SUSY masses:
$m_{\tilde{e}_L} \approx m_{\tilde{e}_R} > m_{\widetilde{\chi}_1^0}$.}
Thus, the final state consists of a charged-lepton pair and a missing
energy. We can construct the cluster transverse mass given by
\begin{eqnarray}
M_T &=& 
 \left[ \left( \sqrt{p_{Te^+ e^-}^2 + M_{e^+ e^-}^2} + \not\!{p}_T \right )^2
  - \left( \vec{p}_{Te^+ e^-} + \vec{\not\!{p}}_T \right )^2 \right ]^{1/2} 
  \nonumber \\
 & =& p_{Te^+e^-} + \sqrt{p_{Te^+ e^-}^2 + M_{e^+ e^-}^2} \label{cluster} \;,
\end{eqnarray}
where the second equality is because $\vec{p}_{Te^+e^-} = - \vec{\not\!{p}}_T$.
We show the distribution for this cluster transverse mass in Fig.~\ref{mt}.
We have imposed a set of leptonic cuts before we construct the
cluster transverse mass:
\begin{equation}
\label{leptoncut}
p_{Te} > 30\;{\rm GeV},\;\; |\eta_e|< 3,\;\; \not\!{p}_T > 50\;{\rm GeV} \;.
\end{equation}
The $Z'$ models shown in Fig.~\ref{mt} are $Z'_N$, $Z'_\eta$, $Z'_\chi$,
and $Z'_{B-L}$. 
The other $Z'$ models show similar features.
The underneath curve is the MSSM contribution only with $\gamma$ and $Z$
exchanges while the upper curve includes also the contribution from $Z'$.
The $Z'$ peak becomes broad because of the missing energies from the two
neutralinos involved. Nevertheless, the sharp edge of the peak is sensitive
to the mass difference between the $Z'$ and the slepton masses.

\subsection{Neutralino-pair production}

Next, we study the production of a neutralino pair $\widetilde{\chi}^0_1 
\widetilde{\chi}^0_2$.
Assuming $\widetilde{\chi}^0_1$ is the LSP, then
$\widetilde{\chi}^0_2$ can decay into $\widetilde{\chi}^0_1 \ell^+ \ell^-$
via a virtual $Z$ boson or a virtual slepton.  The final state consists
of a charged lepton pair and a missing energy.  We can again construct
the cluster transverse mass as in Eq.~(\ref{cluster}).

Electroweak production of $\tilde{\chi}_1^0 \tilde{\chi}_2^0$
goes through the $Z$ and $Z'$ exchanges. We assume that the squarks are 
much heavier such that the $t$-channel squark exchanges are suppressed.
The differential cross section for the subprocess
$ q(p_1) \bar{q}(p_2) \to \tilde{\chi}_1^0(k_1) \tilde{\chi}_2^0(k_2) $ 
is given by
\begin{eqnarray}
\frac{d \hat \sigma}{d \cos \hat{\theta}} = 
   \frac{\beta^\prime}{96 \pi \hat{s}}\!\!\!
&\; \bigg\{&\!\!
( m_{\tilde{\chi}_1^0}^2 - \hat{u} )\,(m_{\tilde{\chi}_2^0}^2 - \hat{u} )\,
\left( |M_{LL}(\hat{s})|^2 + |M_{RR}(\hat{s})|^2 \right)
\nonumber  \\
&+&\!\!
( m_{\tilde{\chi}_1^0}^2 - \hat{t} )\,(m_{\tilde{\chi}_2^0}^2 - \hat{t} )\,
\left( |M_{LR}(\hat{s})|^2 + |M_{RL}(\hat{s})|^2 \right)
\nonumber \\
&+& \!\!
2\,m_{\tilde{\chi}_1^0}\,m_{\tilde{\chi}_2^0}\,\hat{s}\,
\Big(\,\text{Re}\,\big[ M_{LL}(\hat{s})\,M_{RL}^*(\hat{s}) \big] 
\nonumber \\
&& + 
\text{Re}\,\big[ M_{RR}(\hat{s})\,M_{LR}^*(\hat{s}) \big]\Big)
\bigg\}\; ,
\end{eqnarray}
where 
\begin{eqnarray}
M_{\alpha\beta}(\hat s) &=&   
   \frac{g_1^2}{ \hat s - m_Z^2 + i m_Z \Gamma_Z} O^{''\alpha}_{12} g^{q}_\beta
 + \frac{g_2^2}{ \hat s - m_{Z'}^2 + i m_{Z'} \Gamma_{Z'}} C^{''\alpha}_{12} 
 Q'^{q}_\beta \; ,
\end{eqnarray}
$\alpha, \beta = L, R$ and
\begin{equation}
\beta^{\prime} = \Big\{ \big[ \, 1 - (m_{\tilde{\chi}_1^0}^2 
+ m_{\tilde{\chi}_2^0}^2) / \hat s \, \big]^2  
- (\, 2\,m_{\tilde{\chi}_1^0}\,m_{\tilde{\chi}_2^0}/ \hat s \, )^2 \Big\} ^{1/2}\, .  
\end{equation}\\
Here the chiral couplings $O^{''\alpha}_{12}$ of the $Z$ boson and the 
chiral couplings $C^{''\alpha}_{12}$ of the $Z'$ boson to the neutralinos are, 
respectively, given by
\begin{eqnarray}
O^{''L}_{12} &=& -\frac{1}{2}N_{13}N^*_{23} + \frac{1}{2}N_{14}N^*_{24} \; ,
\nonumber  \\
O^{''R}_{12} &=& -O^{''L*}_{12}\;, 
\end{eqnarray}
and
\begin{eqnarray}
C^{''L}_{12} &=& -Q'_{H_d}N_{13}N^*_{23} - Q'_{H_u}N_{14}N^*_{24} \; ,
\nonumber  \\
C^{''R}_{12} &=& -C^{''L*}_{12}\;,
\end{eqnarray}
where $N$ is the mixing matrix of the neutralinos defined in the appendix.
Numerically,
with the choice of SUSY parameters of set (A), we obtain 
the masses
$m_{\tilde{\chi}_1^0}=74.8$ GeV  and $m_{\tilde{\chi}_2^0} = 130.6$ GeV,
and the mixing parameters 
$N_{13}=0.52$, $N_{23}=-0.41$, $N_{14}=-0.33$, and $N_{24}=0.38$.

\begin{figure}[h!]
\centering
\includegraphics[width=5in]{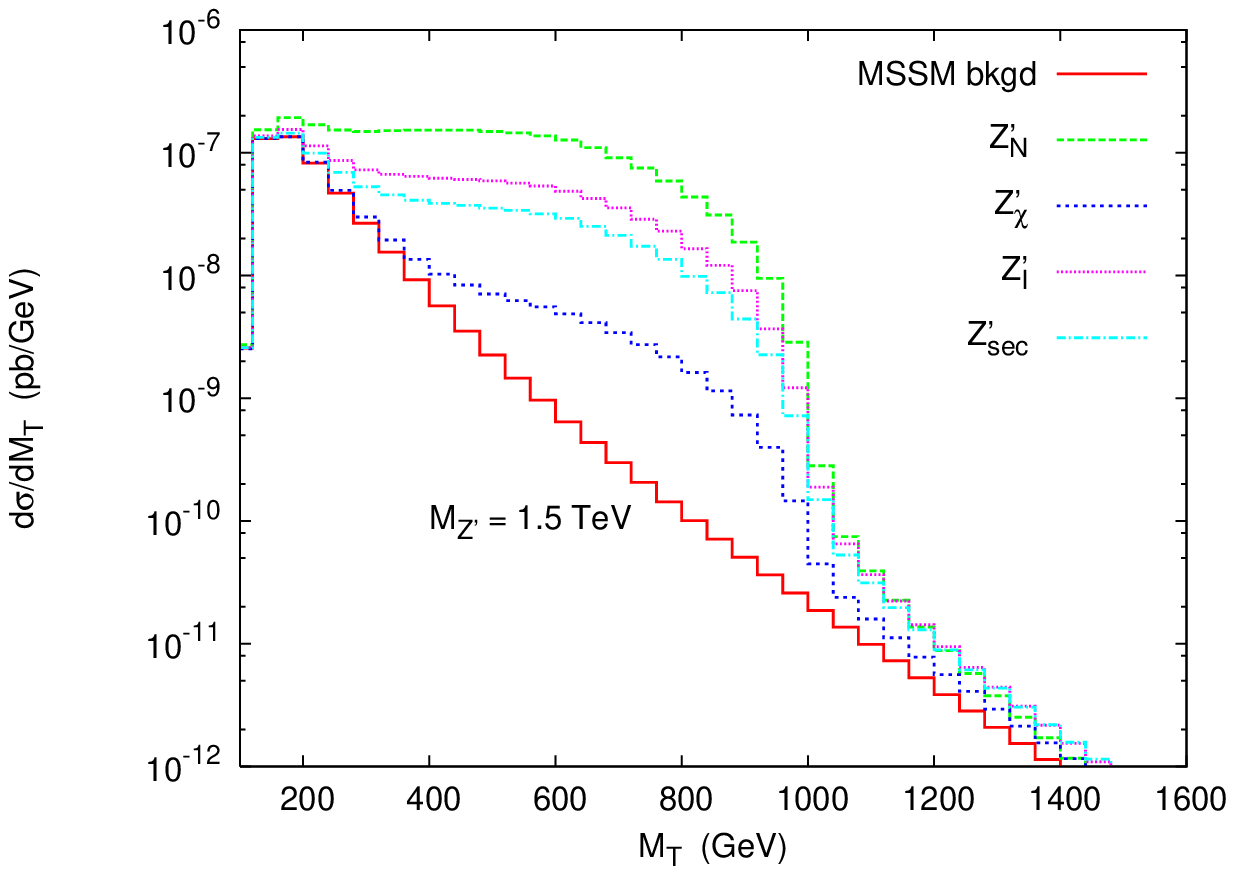}
\medskip
\caption{\small \label{mt-2}
The differential cross section versus the cluster transverse mass $M_T$
defined in Eq.~(\ref{cluster}) for 
$pp \to Z, Z' \to \widetilde{\chi}^0_1 \widetilde{\chi}^0_2$ followed
by the leptonic decay of $ \widetilde{\chi}^0_2 \to \widetilde{\chi}^0_1
e^+ e^-$.
We have applied the leptonic cuts given in Eq.~(\ref{leptoncut})
and assumed the branching ratio 
$B(\widetilde{\chi}^0_2 \to \widetilde{\chi}^0_1 e^+ e^-) = 0.1$.
The underneath curve
is without the $Z'$ boson while the upper curve includes the $Z'$ boson.
The $Z'$ models shown here are $Z'_N$, $Z'_\chi$, $Z'_I$,
and $Z'_{\rm sec}$.
The masses are $M_{Z'}=1.5$ TeV, $m_{\tilde{\chi}_1^0}=74.8$ GeV,
and $m_{\tilde{\chi}_2^0} = 130.6$ GeV.}
\end{figure}

We apply the same set of leptonic cuts as in Eq.~(\ref{leptoncut})
and construct the cluster transverse mass.  We show the cluster
transverse-mass spectrum in Fig.~\ref{mt-2}. 
The $Z'$ models shown in Fig.~\ref{mt-2} are $Z'_N$, $Z'_\chi$, $Z'_I$,
and $Z'_{\rm sec}$. 
The underneath curve is the MSSM contribution only with the $Z$
exchange while the upper curve includes also the contribution from $Z'$.
The $Z'$ peak becomes broad because of the missing energies from the two
neutralinos involved. Nevertheless, the edge of the peak is sensitive
to the mass difference between the $Z'$ and the neutralinos.

\subsection{Chargino-pair production}

Lastly, we have electroweak production of $\tilde{\chi}_1^+ \tilde{\chi}_1^-$
that goes through the $\gamma, \,Z,$ and $Z'$ exchanges.
The differential cross section for the subprocess
$ q(p_1) \bar{q}(p_2) \to \tilde{\chi}_1^+(k_1) \tilde{\chi}_1^-(k_2) $ is given by
\begin{eqnarray}
\frac{d \hat \sigma}{d \cos \hat{\theta}} = \frac{\beta}{96 \pi \hat{s}}\!\!\!
&\;\bigg\{&\!\!
( m_{\tilde{\chi}_1^+}^2 - \hat{u} )^2 \,
\left( |M_{LL}(\hat{s})|^2 + |M_{RR}(\hat{s})|^2 \right)
\nonumber  \\
&+&\!\!
( m_{\tilde{\chi}_1^+}^2 - \hat{t} )^2 \,
\left( |M_{LR}(\hat{s})|^2 + |M_{RL}(\hat{s})|^2 \right)
\nonumber \\
&+& \!\!
2\,m_{\tilde{\chi}_1^+}^2\,\hat{s}\,
\Big(\,\text{Re}\,\big[ M_{LL}(\hat{s})\,M_{RL}^*(\hat{s}) \big] 
\nonumber \\
&& + 
\text{Re}\,\big[ M_{RR}(\hat{s})\,M_{LR}^*(\hat{s}) \big]\Big)
\bigg\}\; ,
\end{eqnarray}
where 
\begin{eqnarray}
M_{\alpha\beta}(\hat s) &=&\!\! 
 \frac{e^2 Q_{\tilde{\chi}^+} Q_q}{\hat s} 
 + \frac{g_1^2 O^{'\alpha}_{11} g^{q}_\beta}{ \hat s - m_Z^2 + i m_Z \Gamma_Z } 
 + \frac{g_2^2 C^{'\alpha}_{11} Q'^{q}_\beta}{ \hat s - m_{Z'}^2 + i m_{Z'} \Gamma_{Z'} } \; ,
\end{eqnarray}
$\alpha, \beta = L, R$\,, and
$\beta = \sqrt{1 - 4  m_{\tilde{\chi}_1^+}^2/ \hat s}$.
Here the chiral couplings $O^{'\alpha}_{11}$ of the $Z$ boson and 
the couplings $C^{'\alpha}_{11}$ of the $Z'$ boson to charginos are, 
respectively,
\begin{eqnarray}
O^{'L}_{11} &=& |V_{11}|^2 + \frac{1}{2}|V_{12}|^2 - x_{\rm w}\;,
\nonumber  \\
O^{'R}_{11} &=& |U_{11}|^2 + \frac{1}{2}|U_{12}|^2 - x_{\rm w}\;, 
\end{eqnarray}
and
\begin{eqnarray}
C^{'L}_{11} &=& Q'_{H_u} |V_{12}|^2\;,
\nonumber  \\
C^{'R}_{11} &=& -Q'_{H_d} |U_{12}|^2\;,
\end{eqnarray}
where $U$ and $V$ are the mixing matrices of the charginos.
Numerically, with the choice in set (A) for SUSY parameters 
we have $V_{12}=0.75$, $V_{11}=-0.67$,
$U_{12}=0.89$, $U_{11}=-0.46$ and $m_{\tilde{\chi}_1^+}=109.5$ GeV.

\begin{figure}
\centering
\includegraphics[width=5in]{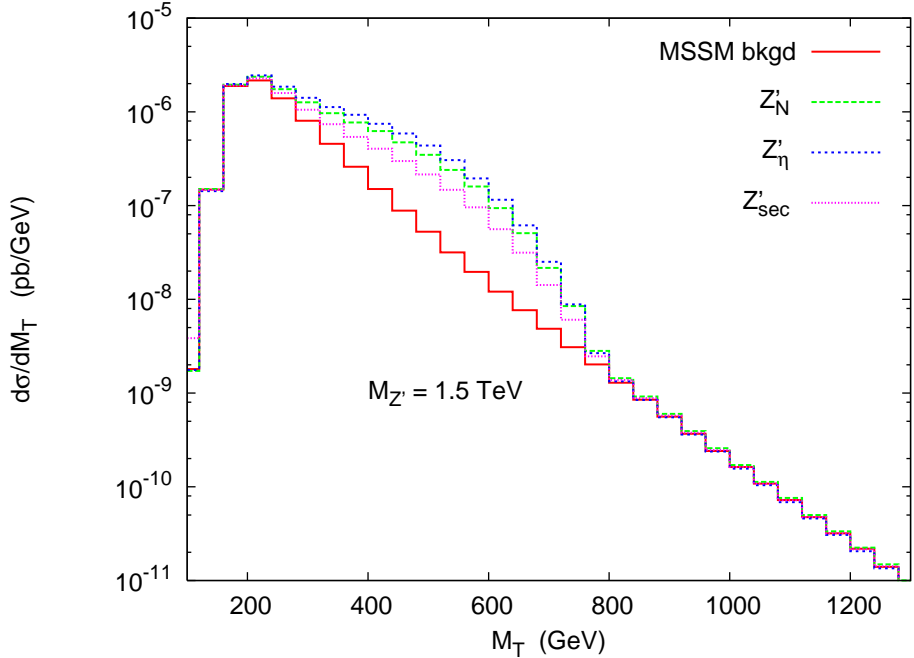}
\caption{\small \label{chargino}
The differential cross section versus the cluster transverse mass $M_T$
defined in Eq.~(\ref{cluster}) for 
$pp \to \widetilde{\chi}^+_1 \widetilde{\chi}^-_1$ followed
by the leptonic decay of $ \widetilde{\chi}^+_1 \to \widetilde{\chi}^0_1
e^+ \nu_e$.
We have applied the leptonic cuts given in Eq.~(\ref{leptoncut})
and assumed the branching ratio 
$B(\widetilde{\chi}^+_1 \to \widetilde{\chi}^0_1 e^+ \nu_e)
\times B(\widetilde{\chi}^-_1 \to \widetilde{\chi}^0_1 e^- \bar \nu_e)
 = 0.1$.
The underneath curve
is without the $Z'$ boson while the upper curve includes the $Z'$ boson.
The $Z'$ models shown here are $Z'_N$, $Z'_\eta$,
and $Z'_{\rm sec}$.
The masses are $M_{Z'} =1.5$ TeV and $m_{\tilde{\chi}_1^+}=109.5$ GeV.} 
\end{figure}

We only calculate the production of $\widetilde{\chi}^+_1 \widetilde{\chi}^-_1$,
because the second chargino is about twice as heavy as the first one.
Each of the charginos decays via a virtual $W$, $\tilde{\nu}$, or 
$\tilde{\ell}$ into a charged lepton, a neutrino, and the lightest neutralino
(if going through the virtual $W$, light quarks are also possible).
Therefore, there will two charged leptons plus missing energies in the
final state.  Just as the same as the case of slepton-pair or neutralino-pair
production, we reconstruct the cluster transverse mass as in 
Eq.~(\ref{cluster}).
We show the distribution of cluster transverse mass in Fig.~\ref{chargino}
for $Z'_N$, $Z'_\eta$, and $Z'_{\rm sec}$.
It is easy to see the bump due to the presence of the $Z'$ boson,
though the bump is not as discernible as the previous two 
cases of slepton-pair and neutralino-pair production.

\section{Conclusions}
In this work, we have studied the possible supersymmetric decay modes of
an additional neutral gauge boson $Z'$, which is currently limited to be at least 1 TeV.
Grand unified theories have predicted one or more such $Z'$ bosons 
along the path through which the GUT symmetry is broken down to the 
electroweak symmetry.  When supersymmetry is included in the theory,
such a $Z'$ boson can decay not only into the SM particles but also
the supersymmetric partners.  We have used $E_6$, $U(1)_{B-L}$, and the
sequential models to illustrate how the decays of the $Z'$ are affected.
In particular, the golden search mode--charged leptons-- for the $Z'$ 
will have a smaller branching ratio as the supersymmetric modes open. 
We have shown that the current limits obtained at the Tevatron and the
LHC will be reduced by a noticeable amount, of order 20 GeV. 
We have also estimated the 5$\sigma$ discovery sensitivities 
of the $Z'$ at the LHC, including the effect of supersymmetric decay modes.
Finally, we demonstrated that even though the $Z'$ decays into 
supersymmetric particles, giving rise to missing energies, one can still
reconstruct the cluster transverse mass (using the observable charged leptons)
to identify the existence of the $Z'$.
We believe further studies along this direction is worthwhile since it 
can help us to fully understand
the role of the $Z'$ in the supersymmetry-breaking and the symmetry 
breaking pattern.

\section*{Acknowledgments}
The work was supported in parts by the National Science Council of
Taiwan under Grant Nos. 99-2112-M-007-005-MY3 and
98-2112-M-001-014-MY3, and the WCU program through the KOSEF
funded by the MEST (R31-2008-000-10057-0).


\appendix

\section{Feynman Rules}

As explained in Sec. IIB, the extra term $h_s S H_u H_d$ allowed by 
the $U(1)'$ symmetry will generate the effective $\mu$ term when the
singlet scalar field $S$ develops a VEV. With the assumption that 
the singlet scalar and singlino fields, $Z'$-ino, and other exotic
fermions are very heavy, the low-energy particle content includes
the $Z'$ boson and those of MSSM.
The effective superpotential $W_{\rm eff}$ involving the matter and Higgs 
superfields is the same as the MSSM's, and given by 
\begin{equation}
W_{\rm eff} = \epsilon_{ab} \left [ y^u_{ij} Q^a_j H_u^b U^c_i 
  - y^d_{ij} Q^a_j H_d^b D^c_i 
  -  y^l_{ij} L^a_j H_d^b E^c_i
  +  \mu  H_u^a H_d^b \right ] \;,
\end{equation}
where $\epsilon_{12}= - \,\epsilon_{21} =1$, $i,j$ are family indices,
and $y^u$ and $y^d$ represent the Yukawa matrices for the 
up-type and down-type quarks respectively.
Here $Q, L, U^c, D^c, E^c, H_u$, and $H_d$ denote the superfields for the 
quark doublet, lepton doublet,
up-type quark singlet, down-type quark singlet, lepton singlet,
up-type Higgs doublet, and down-type Higgs doublet respectively.  
The scalar interactions are obtained by calculating the 
$F$- and $D$-terms of the superpotential, and by including the 
following soft-SUSY-breaking terms
\begin{eqnarray}
{\cal L}_{\rm soft} &=& - M_1 \widetilde{B} \widetilde{B} 
                        - M_2 \widetilde{W} \widetilde{W}
                        - M_3   \tilde{g} \tilde{g} \nonumber \\
          & -  & M_{\tilde{Q}}^2 \widetilde{Q}^\dagger \widetilde{Q}
             - M_{\tilde{U}^c}^2 \widetilde{U}^{c\dagger} \widetilde{U}^c
             - M_{\tilde{D}^c}^2 \widetilde{D}^{c\dagger} \widetilde{D}^c
             - M_{\tilde{L}}^2 \widetilde{L}^{\dagger} \widetilde{L}
             - M_{\tilde{E}^c}^2 \widetilde{E}^{c\dagger} \widetilde{E}^c 
  \\
{\cal L}_{\rm tril} &=& 
       \epsilon_{ab} \left [ y^u_{ij} A^U_{ij} 
                  \widetilde{Q}^a_j H_u^b \widetilde{U}^c_i 
  - y^d_{ij} A^D_{ij} \widetilde{Q}^a_j H_d^b \widetilde{D}^c_i 
  -  y^l_{ij} A^E_{ij} \widetilde{L}^a_j H_d^b \widetilde{E}^c_i 
  +  \mu  B H_u^a H_d^b \right ] \;,
\end{eqnarray}
where the ${\cal L}_{\rm soft}$ represents the 
soft mass terms for the gauginos
\footnote{
There should also be a soft gaugino mass term for the $Z'$-ino. 
However, we will assume it is heavy and let it decouple from the 
low energy spectrum in the present work.} 
and sfermions,
and ${\cal L}_{\rm tril}$ represents the trilinear $A$ terms.

The gauge interactions for the fermionic and scalar 
components, denoted generically by $\psi$ and $\phi$ respectively,  
of each superfield mentioned above are given by
\begin{equation}
\label{a1}
 {\cal L} = \bar  \psi \, i \gamma^\mu D_\mu \, \psi 
   + ( D^\mu \phi)^\dagger \, (D_\mu \phi) \;,
\end{equation}
where the covariant derivative is defined as usual
\begin{equation}
\label{a2}
  D_\mu = \partial_\mu + i e Q A_\mu + i \frac{g }{\sqrt{2}} (\tau^+ W^+_\mu
  +\tau^- W^-_\mu ) + i g_1 ( T_{3L} - x_{\rm w} Q ) Z_\mu 
        + i g_2  Z'_\mu  Q'  \;.
\end{equation}
Here $g_1 = g / \cos \theta_{\rm w}$ with $g$ the $SU(2)_L$ gauge coupling 
and  $\theta_{\rm w}$ the weak mixing angle, $g_2$ is the gauge 
coupling for the extra $U(1)$,
$\tau^\pm$ and $T_{3L}$ are the ladder operators 
and the third component of the $SU(2)_L$ generators,
$Q$ and $Q'$ are the charges of the two $U(1)$s 
in unit of the electromagnetic charge $e$ and $g_2$ respectively, 
and finally, $x_{\rm w} = \sin^2 \theta_{\rm w}$.
Since we only consider supersymmetric $U(1)$ symmetry for the $Z'$ boson 
and due to the Majorana nature of the $Z'$-ino,
there is no coupling between the $Z'$-ino and the $Z'$ boson.
Simply from Eqs.~(\ref{a1}) and (\ref{a2}) we obtain the 
interactions of $Z'$ with fermions, sfermions, neutral and charged
Higgsinos (which become the physical neutralinos and charginos after mixing 
effects are taken into account),
and the Higgs bosons.  

The rotation of neutral bino, wino, and Higgsinos into the physical 
neutralinos is given by
\begin{equation}
  \left( \begin{array}{c} 
        \widetilde{\chi}^0_1 \\
        \widetilde{\chi}^0_2 \\
        \widetilde{\chi}^0_3 \\
        \widetilde{\chi}^0_4  \end{array} \right ) = N\;
\left( \begin{array}{c} 
        \widetilde{B} \\
        \widetilde{W}^3 \\
        \widetilde{H}_d^0 \\
        \widetilde{H}_u^0  \end{array} \right ) \; ,
\end{equation}
where $N$ is an orthogonal matrix.
The rotation of the charged wino and Higgsino into the physical charginos is 
via bi-unitary transformation
\begin{equation}
  \left( \begin{array}{c} 
 \widetilde{\chi}^+_1 \\
 \widetilde{\chi}^+_2 \end{array} \right ) = V\, 
   \left( \begin{array}{c} 
 \widetilde{W}^+ \\
 \widetilde{h}^+_u \end{array} \right )  \;, \qquad \qquad
  \left( \begin{array}{c} 
 \widetilde{\chi}^-_1 \\
 \widetilde{\chi}^-_2 \end{array} \right ) = U\, 
   \left( \begin{array}{c} 
 \widetilde{W}^- \\
 \widetilde{h}^-_d \end{array} \right )  \; ,
\end{equation}
where $U$ and $V$ are unitary matrices.
The mixing angles involved in physical Higgs bosons ($h^0, \, H^0, \, H^\pm,\,
A^0$) can be read off from the following decompositions of the 
Higgs boson fields $H_u$ and $H_d$:
\begin{equation}
 H_u =  \left( \begin{array}{c}
                      h_u^+ \\
                      h_u^0  \end{array}   \right )
   = \left ( \begin{array}{c}
   H^+ \cos\beta  + G^+ \sin\beta  \\
  \frac{v\sin\beta}{\sqrt{2}} + \frac{1}{\sqrt{2}}(h^0 \cos\alpha + 
 H^0 \sin\alpha ) + i (A^0  \cos\beta + G^0 \sin\beta ) 
       \end{array}  \right ) \; ,
\end{equation}
\begin{equation}
 H_d = \left( \begin{array}{c}
                      h_d^0 \\
                      h_d^-  \end{array}   \right )
= \left( \begin{array}{c}
  \frac{v\cos\beta}{\sqrt{2}} + \frac{1}{\sqrt{2}}(- h^0 \sin\alpha + 
 H^0 \cos\alpha ) + i ( A^0 \sin\beta  - G^0 \cos\beta )  \\
     H^- \sin\beta - G^- \cos\beta   
         \end{array}
  \right ) \;,
\end{equation}
where the angle $\alpha$ is the mixing of the neutral CP-even Higgs bosons
$h^0$ and $H^0$, $G^{0,\pm}$ are the Goldstone bosons, and $v=246$ GeV.

Lastly, there are Yukawa-type interactions between the gauginos and
the scalar $\phi$ and fermionic $\psi$ components of the matter superfield
of the following form
\begin{equation}
{\cal L} = - \sqrt{2} g^a \phi^* \,T^a \,\widetilde{\lambda}^a \, \psi 
     +{\rm h.c.}
\end{equation}
where $a$ is the group index of the $U(1)_Y$, $SU(2)_L$, $SU(3)_C$, or
the extra $U(1)_{Z'}$. The interactions involving the $Z'$-ino are
\begin{eqnarray}
 {\cal L}_{\tilde{Z}'} &= & -\sqrt{2} g_2  \Biggr [ 
  \widetilde{Q}^\dagger \widetilde{Z'}  Q'_{Q} Q 
+  \widetilde{U}^{c \dagger} \widetilde{Z'}  Q'_{U^c} U^c
+  \widetilde{D}^{c \dagger} \widetilde{Z'}  Q'_{D^c} D^c  + {\rm h.c.} 
            \nonumber \\
&& +  \widetilde{L}^\dagger \widetilde{Z'}  Q'_{L} L 
+  \widetilde{E}^{c\dagger} \widetilde{Z'}  Q'_{E^c} E^c  + {\rm h.c.} 
        \nonumber \\
&& +  H_u^\dagger \widetilde{Z'}  Q'_{H_u} \widetilde{H}_u  
+  H_d^\dagger \widetilde{Z'}  Q'_{H_d} \widetilde{H}_d   + {\rm h.c.}   \Biggr ]
\;.
\end{eqnarray}
Note that the $Z'$-ino will mix with the $\widetilde{H}_u$ and 
$\widetilde{H}_d$
Higgsinos when the $H_u^0$ and $H_d^0$ take on vacuum expectation values.
Thus, we will have a $5\times 5 $ neutralino mass matrix. 
However, we decouple the $Z'$-ino in this work by setting the $Z'$-ino mass
heavy. We will come back to this in later work.

In the following we list the Feynman diagrams and the corresponding
Feynman rules involving the $Z'$ boson and the MSSM particles. 
For each model, one should use the corresponding coupling strength 
and chiral charges. 
The coupling strength $g_2$ and 
the $Z'$ charges are given in Tables~\ref{E6} and \ref{b-l} for 
the $E_6$ and $U(1)_{B-L}$ models respectively.
For the sequential $Z'$ model, replace the coupling strength $g_2$ by
$g_1$ while the corresponding chiral charges are given by the SM values:
$Q'_i (Z'_{SM})= T_{3i} - x_{\rm w} Q_i$.
The chiral couplings of the charginos and neutralinos with the $Z'$ boson 
are, respectively, given by
\begin{eqnarray}
&& C^{'L}_{ij}=Q'_{H_u} V_{i2} V^*_{j2}\;, \qquad 
             C^{'R}_{ij}=-Q'_{H_d} U^*_{i2} U_{j2} \; , 
 \nonumber \\
&& C^{''L}_{ij} = -Q'_{H_d}N_{i3}N^*_{j3} - Q'_{H_u}N_{i4}N^*_{j4}\;, \qquad
C^{''R}_{ij} = -C^{''L*}_{ij} \; .
\end{eqnarray}
These coupling coefficients are the same for the three $Z'$ models 
that we have studied in this work,
as long as we use the corresponding $Q'$ charges for 
the two Higgs doublet fields.

\begin{itemize}

\item[]
\includegraphics[width=2in]{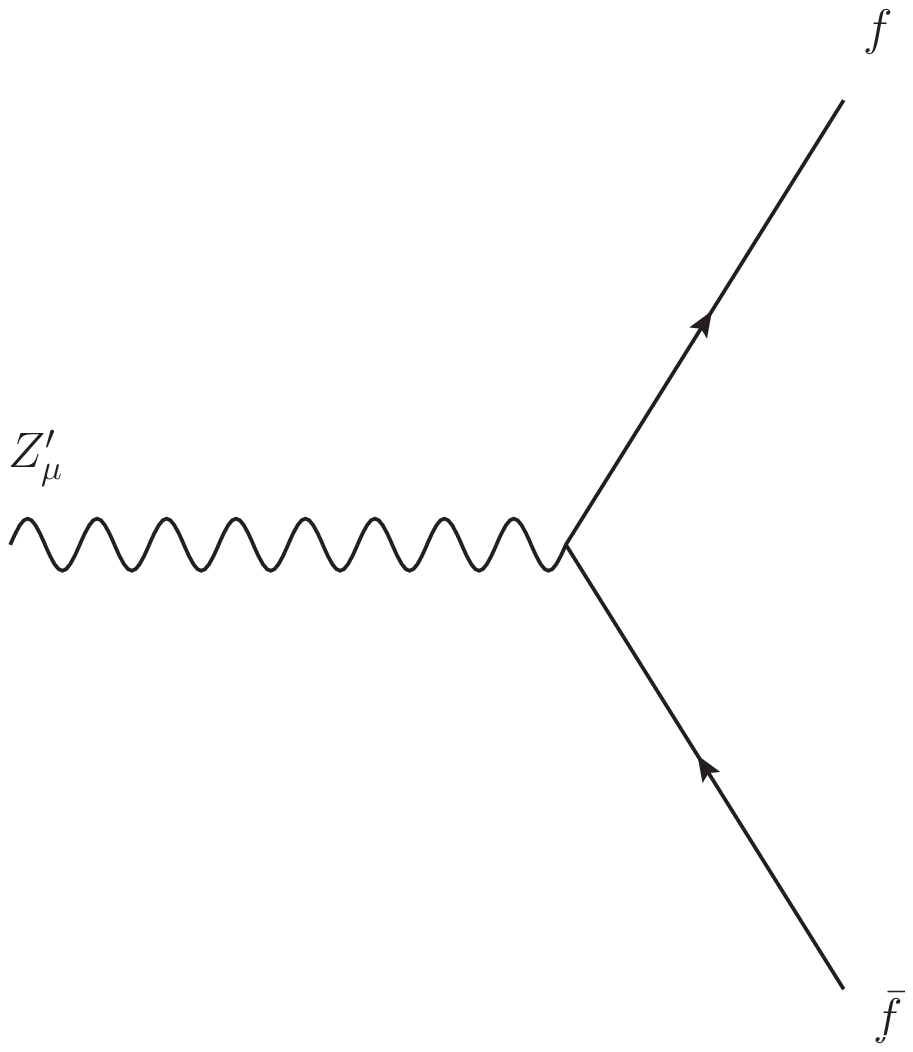}    $\qquad \qquad 
       -i g_2 \gamma^\mu (Q'_{f_L} P_L  + Q'_{f_R} P_R )$ 

\item[]
\includegraphics[width=2in]{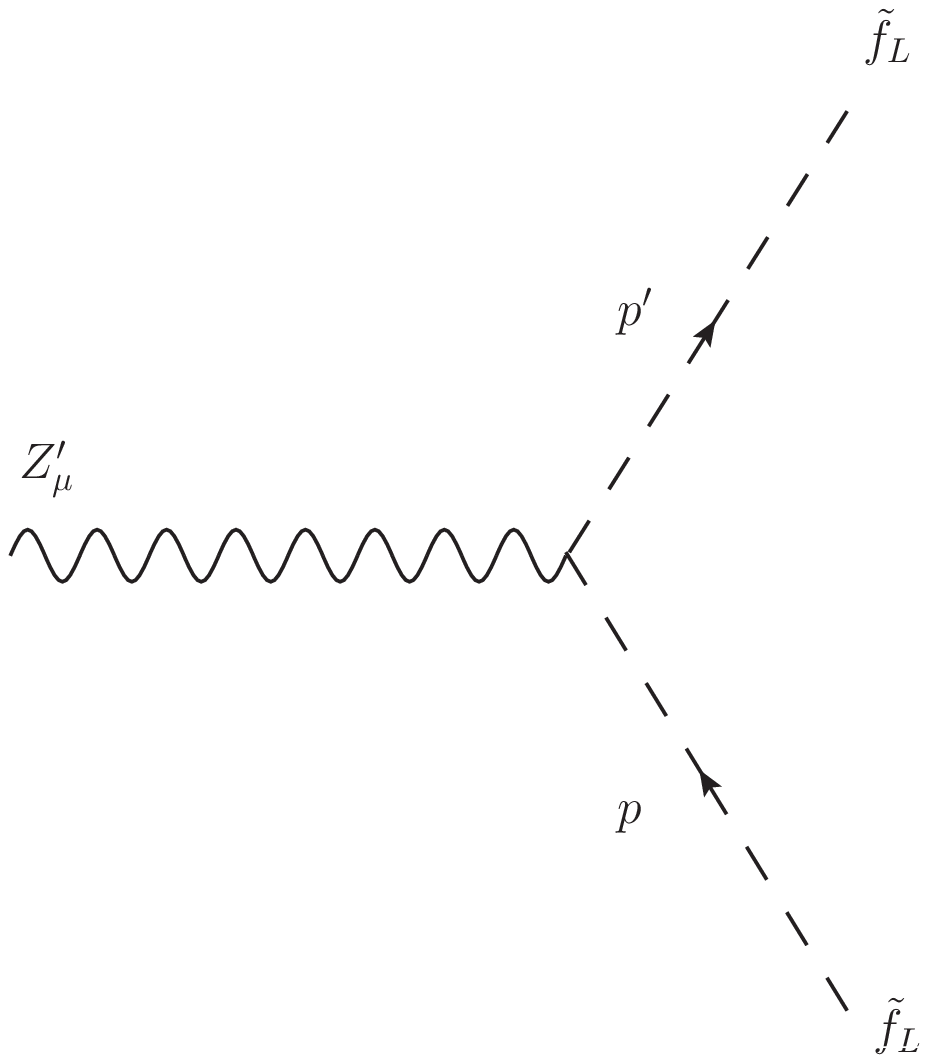}      $\qquad \qquad 
 -i g_2  Q'_{f_L}   (p + p')^\mu$ 
 
\item[]
\includegraphics[width=2in]{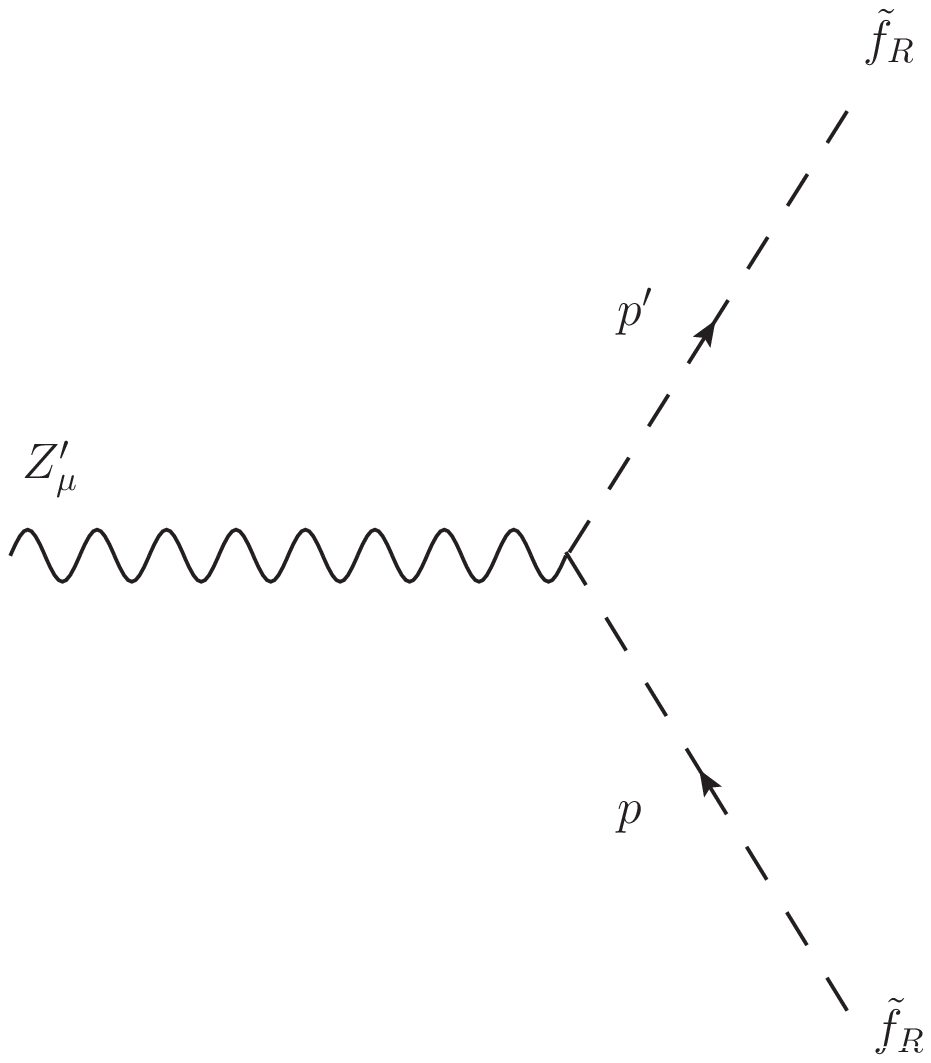}    $\qquad \qquad 
 -i g_2  Q'_{f_R}   (p + p')^\mu$

\item[]
\includegraphics[width=2in]{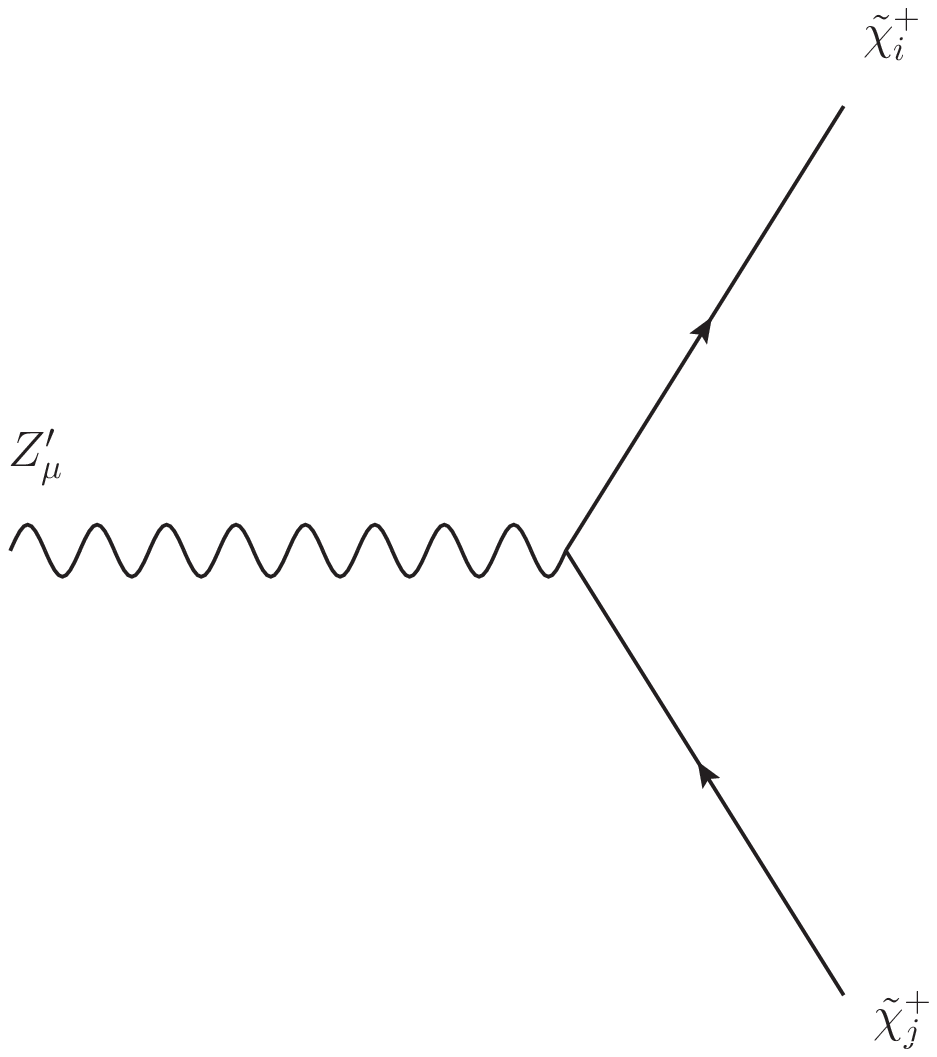}   $\qquad \qquad 
 -i g_2 \gamma^\mu ( C^{'L}_{ij} P_L + C^{'R}_{ij} P_R)$

\item[]
\includegraphics[width=2in]{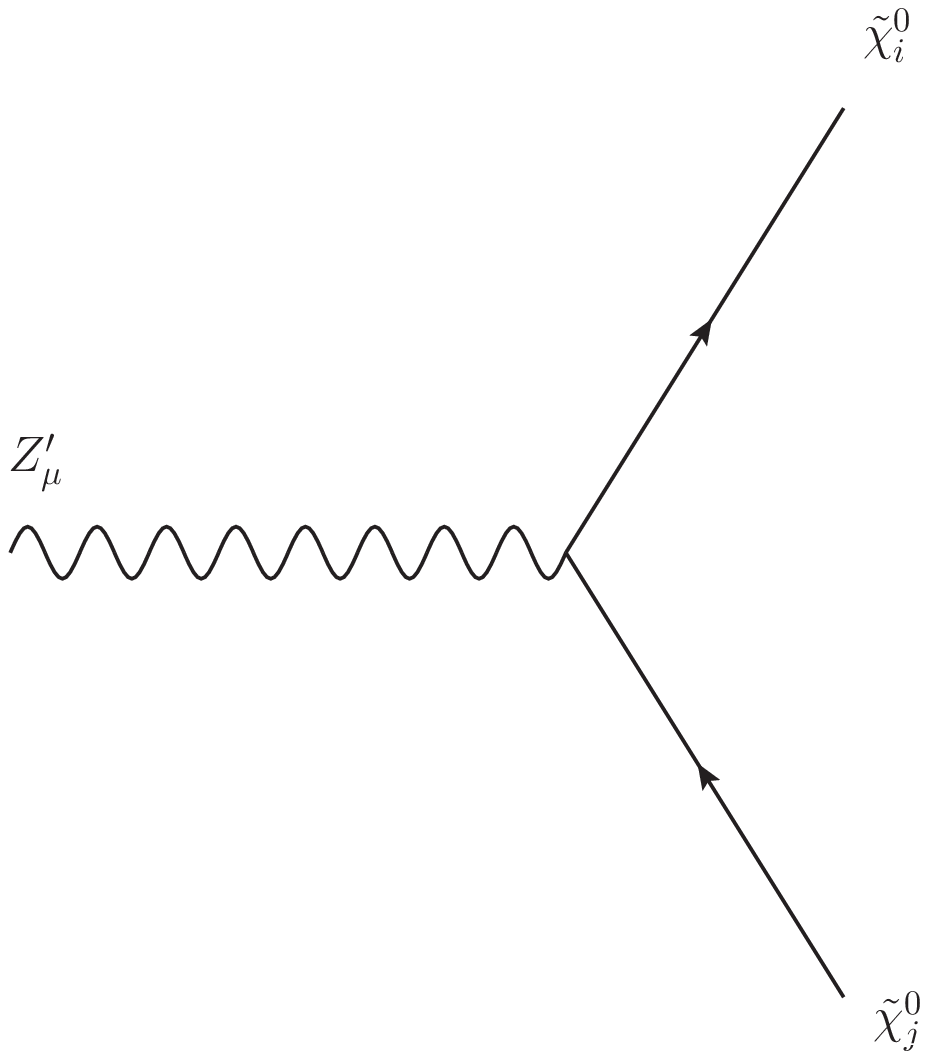}   $\qquad \qquad 
 i g_2 \gamma^\mu ( C^{''L}_{ij} P_L + C^{''R}_{ij} P_R)$

\item[]
\includegraphics[width=2in]{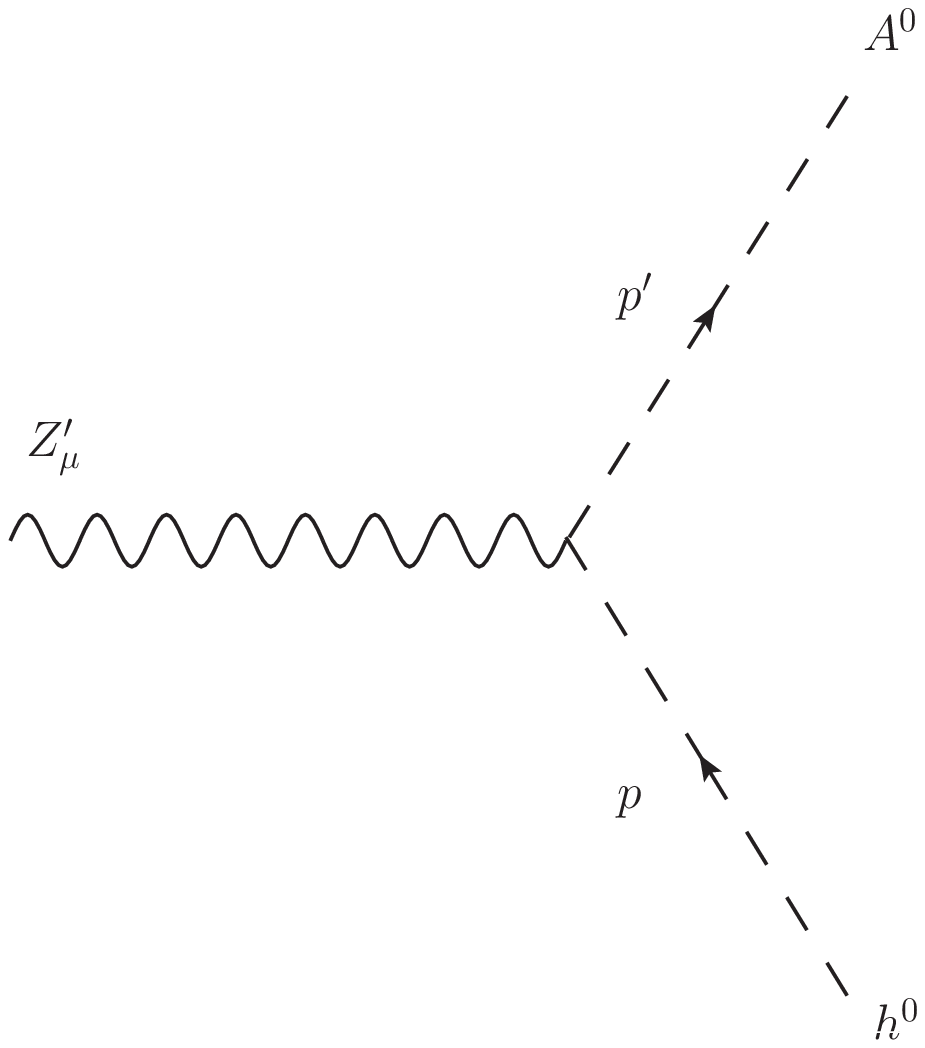}   $\qquad \qquad 
-g_2 (Q'_{H_u} \cos\alpha \cos\beta - Q'_{H_d} \sin\alpha \sin\beta )\,
 (p + p')^\mu$

\item[]
\includegraphics[width=2in]{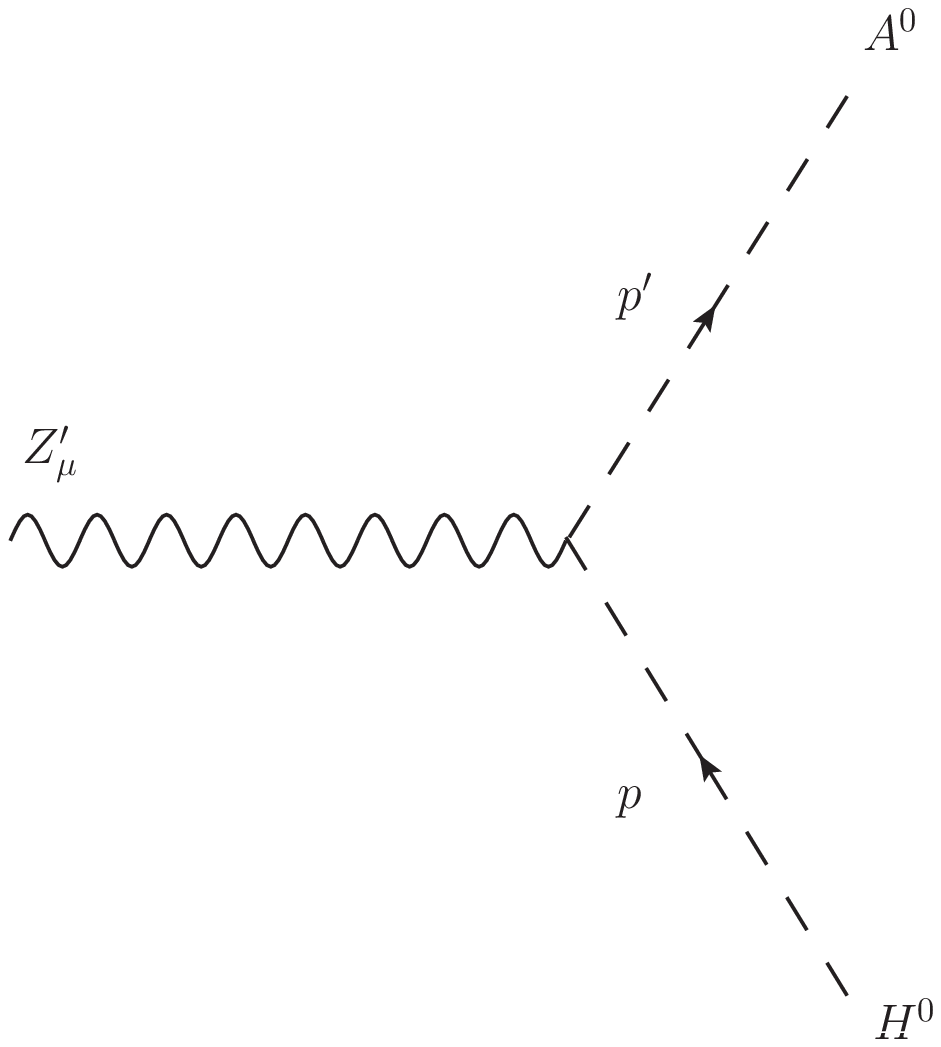}  $\qquad \qquad 
-g_2 (Q'_{H_u} \sin\alpha \cos\beta + Q'_{H_d} \cos\alpha \sin\beta )\,
 (p + p')^\mu$
 
\item[]
\includegraphics[width=2in]{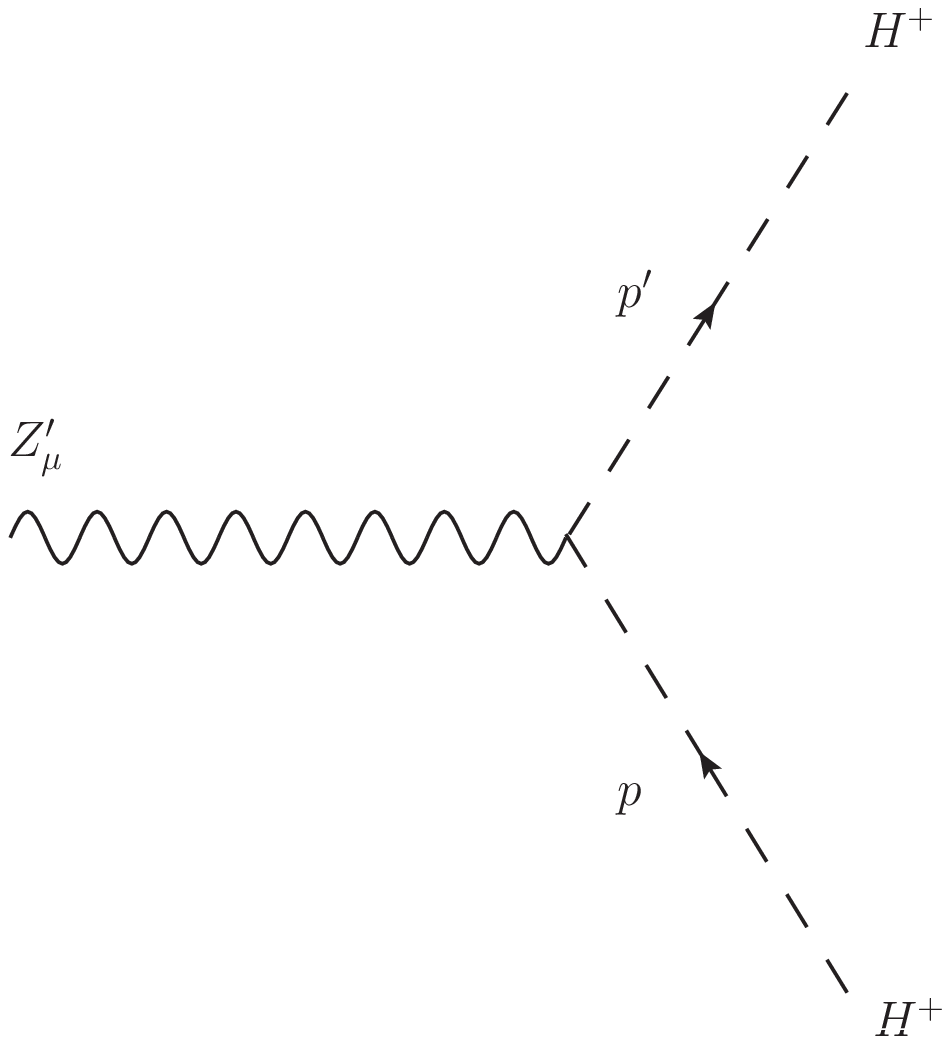}  $\qquad \qquad 
i g_2 (Q'_{H_d} \sin^2\beta - Q'_{H_u} \cos^2\beta )\,
 (p + p')^\mu$ 

\end{itemize}

\end{document}